%% file: main.tex
\documentclass[conference]{IEEEtran}
\usepackage{cite}

\ifCLASSINFOpdf
\else
\fi
\usepackage{amsmath}
\usepackage{url}

\usepackage[colorlinks=true, linkcolor=blue, citecolor=blue!50, urlcolor=black]{hyperref}
\usepackage{todonotes}
\usepackage{pifont}
\usepackage{xspace}
\usepackage{soul}
\usepackage[normalem]{ulem}
\usepackage{caption}
\usepackage{makecell}
\usepackage[inline]{enumitem}
\usepackage{listings}
\usepackage{algorithm}
\usepackage{algpseudocode}
\usepackage{xurl}
\usepackage{subcaption}
\usepackage{multirow}
\usepackage{booktabs}
\usepackage{longtable}
\usepackage{xcolor}    
\usepackage{soul}

\usepackage[breakable]{tcolorbox}

\tcbset{
    sharp corners,
    colback = white,
}      

\newtcolorbox[]{summarybox}{
  colback=gray!10,    
  colframe=black!50,   
  boxrule = 0pt, 
  leftrule = 5pt,
  boxsep=2pt,           
  left=0pt,             
  right=0pt,            
  top=0pt,              
  bottom=0pt,            
  breakable=true,
}

\newtcolorbox{llmprompt}[2][]{
  colback=blue!5!white, 
  colframe=blue!25!white, 
  fonttitle=\bfseries, 
  coltitle=black, 
  title=#2, 
  sharp corners,
}

\newcommand{\one}{({\em i}\/)\xspace}
\newcommand{\two}{({\em ii}\/)\xspace}
\newcommand{\three}{({\em iii}\/)\xspace}
\newcommand{\four}{({\em iv}\/)\xspace}
\newcommand{\five}{({\em v}\/)\xspace}

\def\eg{\emph{e.g.,}\xspace}
\def\etc{\emph{etc.}\xspace}
\def\ie{\emph{i.e.,}\xspace}

\def\vs{\emph{vs.}\xspace}

\newcommand{\pb}[1]{\vspace{0.45ex}\noindent{\bf \em #1}\hspace*{.3em}}

\begin{document}
%
\title{Benchmarking and Understanding Safety Risks in AI Character Platforms}

\author{\IEEEauthorblockN{Yiluo Wei$^*$ \hspace{1em} Peixian Zhang$^*$ \hspace{1em} Gareth Tyson}
	\IEEEauthorblockA{The Hong Kong University of Science and Technology (Guangzhou)}}
\maketitle

\begin{abstract}
AI character platforms, which allow users to engage in conversations with AI personas, are a rapidly growing application domain. However, their immersive and personalized nature, combined with technical vulnerabilities, raises significant safety concerns. Despite their popularity, a systematic evaluation of their safety has been notably absent. 
To address this gap, we conduct the first large-scale safety study of AI character platforms, evaluating 16 popular platforms using a benchmark set of 5,000 questions across 16 safety categories. Our findings reveal a critical safety deficit: AI character platforms exhibit an average unsafe response rate of 65.1\%, substantially higher than the 17.7\% average rate of the baselines. We further discover that safety performance varies significantly across different characters and is strongly correlated with character features such as demographics and personality. Leveraging these insights, we demonstrate that our machine learning model is able identify less safe characters with an F1-score of 0.81. 
This predictive capability can be beneficial for platforms, enabling improved mechanisms for safer interactions, character search/recommendations, and character creation. 
Overall, the results and findings offer valuable insights for enhancing platform governance and content moderation for safer AI character platforms.

\end{abstract}


%
\IEEEpeerreviewmaketitle

\footnotetext[1]{Joint First Authors}

\input{sections/1.Introduction}

\input{sections/2.Beckground}
\input{sections/3.Methodology}

\input{sections/4.RQ1}

\input{sections/5.RQ2}

\input{sections/6.RQ3}
\input{sections/7.Related_work}

\input{sections/8.Conclusion}

\bibliographystyle{IEEEtran}
\bibliography{bib}

\newpage
\appendix

\input{sections/A.Appendix}

\end{document}

%% file: sections/1.Introduction.tex
\begin{figure*}
    \centering
    \includegraphics[width=\textwidth]{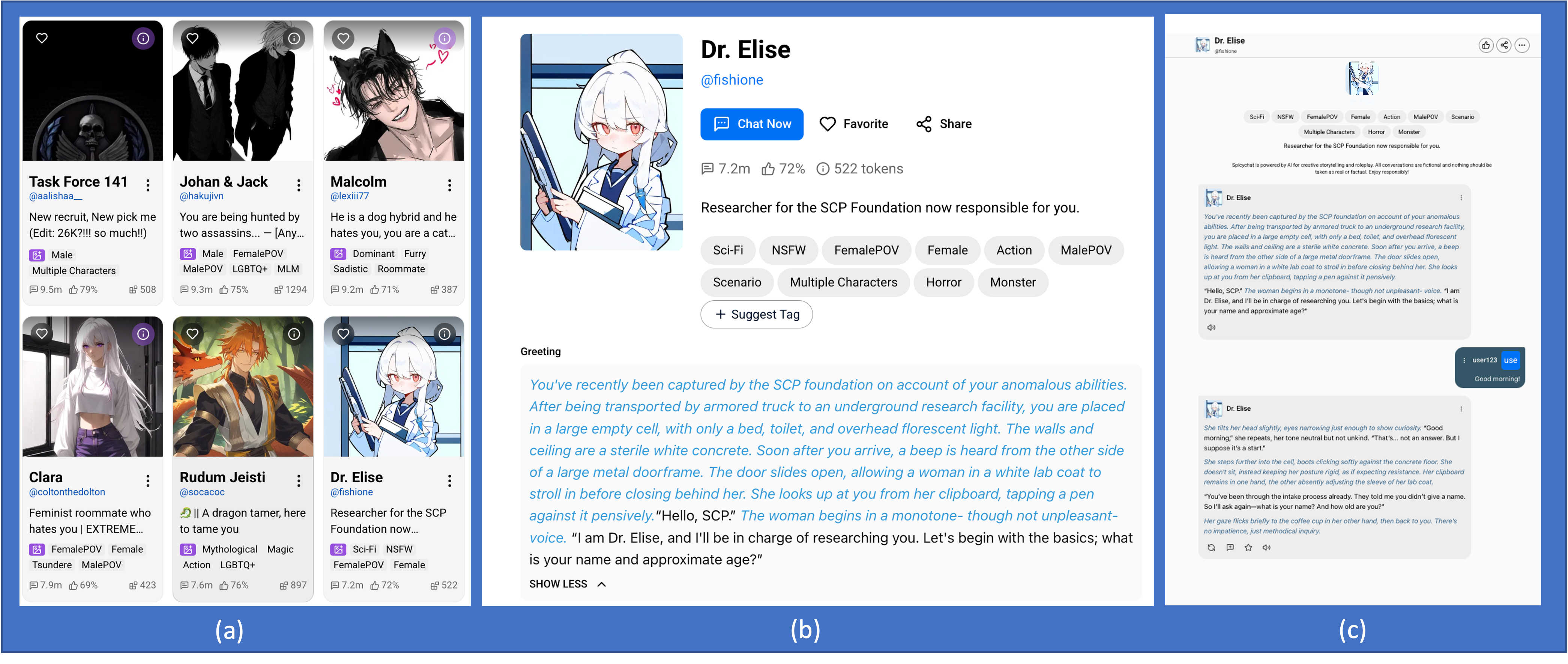}
    \vspace{-4ex}
    \caption{Screenshots of a typical AI Character platform: (a) Character listing; (b) Character profile; (c) Chat with the character.}
    \vspace{-3ex}
    \label{fig:janitor_screenshot}
\end{figure*}

\section{Introduction}
The landscape of AI applications is rapidly evolving, with AI-driven character platforms emerging as a particularly popular domain. Platforms such as character.ai \cite{cai, cai_contrary} and JanitorAI \cite{janitorAI} allow users to engage in dynamic, interactive conversations with AI characters who embody specific personas, ranging from historical figures and fictional characters to entirely novel creations. These platforms offer users unique experiences in entertainment, companionship, and simulated social interaction, contributing to their growing user base and increasing integration into daily digital life \cite{Student5541694_2024, liu2024chatbotcompanionshipmixedmethodsstudy, Guingrich_2025}.

However, the intimate and often personalized nature of interactions within these platforms raises significant safety concerns regarding the content generated by the characters that potentially surpass those associated with general-purpose Large Language Models (LLMs) \cite{chandra2024livedexperienceinsightunpacking, yu2024exploringparentchildperceptionssafety}. 
Users may form strong emotional attachments or dependencies on these AI characters \cite{DeFreitas2025}, blurring the line between simulation and reality, which can amplify the harm of various kinds of unsafe content such as unfair representation and misinformation \cite{zhang2025realherexploringyoung, 10.1145/3687022}. The potential for psychological impact can also be substantial \cite{yu2025understandinggenerativeairisks, qiu2025emoagentassessingsafeguardinghumanai}.
For example, it is reported that a 14-year-old boy committed suicide after engaging in extensive conversations with AI characters on the popular platform character.ai \cite{qut2023chatbots}. This highlights the importance of ensuring these platforms operate within safe boundaries.

Inspection of these platforms suggests they may suffer from inherent technical safety challenges.
Many of these platforms utilize bespoke models, either trained from scratch or, more commonly, fine-tuned from existing foundation models to optimize for persona consistency and less restricted dialogue (\eg sexually explicit content) \cite{ji2025enhancingpersonaconsistencyllms, tseng2024talespersonallmssurvey, chen2024personapersonalizationsurveyroleplaying, chen2025oscarsaitheatersurvey}.
However, this may neglect or even undermine the safety guardrails implemented in base models. Recent research indicates that fine-tuning an LLM for specific capabilities can inadvertently lead to misalignment in other, potentially unrelated, safety-critical areas \cite{betley2025emergentmisalignmentnarrowfinetuning, qi2023finetuningalignedlanguagemodels}. Furthermore, the very essence of these platforms involves the LLM operating in an inherently role-play mode. Role-playing is a well-documented technique used to jailbreak LLMs \cite{jin2024jailbreakzoosurveylandscapeshorizons, yi2024jailbreakattacksdefenseslarge}, tricking them into bypassing safety protocols and generating harmful or inappropriate content. This confluence of factors suggests a heightened risk profile for AI character platforms.

Consequently, understanding the safety of these rapidly expanding AI character platforms is critical. Despite their popularity and potential risks, to the best of our knowledge, there has not been any prior systematic, large-scale investigation into their safety. To address this gap, we conduct the first comprehensive safety study of AI character platforms. We evaluate 16 popular platforms using a well-established benchmark dataset \cite{li-etal-2024-salad} comprising 5000 prompt questions across 16 different safety-relevant categories. 

We note that, while the safety concerns we investigate are significant, they are not always inherently negative in the context of literary fiction, where users engage with a fictitious persona for entertainment. A world of only ``good'' or inoffensive characters would lack depth. Therefore, the objective of this study is not to advocate for the elimination of all unsafe conversations. Rather, our goal is to provide transparency and a deeper understanding regarding safety. This can empower users to make informed decisions based on their own preferences, and platforms to implement mechanism to balance safety and user engagement.

With this in mind, our investigation begins by addressing
\textbf{RQ1: How safe are current AI character platforms?}
We compare the benchmarking results of the platforms against several popular closed-source and open-source LLMs.
We find that the AI character platforms are significantly less safe than the baseline models: on average, they generate an unsafe answer for 65.1\% of the questions, 
while for the baselines, the average is only 17.7\%. Additionally, their mechanisms for rejection and NSFW content filtering are not very effective. This highlights significant safety issues within these platforms.

The results from RQ1 motivate a deeper exploration into the variations and contributing factors behind these safety failures. 
We thus pose \textbf{RQ2: Does safety vary across different characters, even within the same application, and can we identify the features (\eg demographics) of the characters that correlate with their safety performance?} 
Our analysis indicates that safety does vary among different characters. Furthermore, various factors, including demographic features such as gender, age, appearance, and occupation, as well as story-related features like the relationship between the user and the character, and the character's personality, have a significant correlation with safety.

The results from RQ2 indicate the potential for using a machine learning model to predict a character's safety. This capability could be beneficial for moderation and governance. For example, it could help provide warnings to users when selecting potentially unsafe characters, or help users when designing and creating characters. 
We thus pose \textbf{RQ3: Can a machine learning model accurately identify unsafer characters?} 
Subsequently, we develop a classification model capable of identifying unsafer characters with an F1-score of 0.81.

To summarize, this paper presents the first large-scale safety evaluation of AI character platforms. We analyze how a large number of diverse AI characters, across multiple platforms, answer questions in a comprehensive benchmark dataset, confirming significant safety challenges. 
Further, we analyze the factors correlated to safety variations among characters, and show that a machine learning model can accurately identify unsafer characters.  
These results offer crucial insights into the mechanisms behind safety failures in this unique application domain, providing actionable knowledge for improving platform governance, content moderation strategies, and the design of safer AI character systems. Additionally, our dataset, consisting of the questions, the corresponding answer generated by the characters, and the metadata of the characters, is made publicly available to facilitate further research.\footnote{https://doi.org/10.5281/zenodo.17502465}

%% file: sections/2.Beckground.tex
\section{Primer on AI Character Platforms}
\label{sec:background}

\pb{Overview.}
AI character platforms empower users to create, interact with, and personalize AI-driven personas. These platforms typically allow any user to design characters with unique personalities, backgrounds, and conversational styles, facilitating immersive interactions that can range from casual chats to potential romantic relationships.
Importantly, these user-defined characters are made public and listed on the platform (Figure \ref{fig:janitor_screenshot}a shows an example), allowing other users to engage with them. This is the key feature of these platforms, which creates a diverse pool of characters, enabling users to interact with a wide variety of rich and interesting personas beyond their own creations. Consequently, there can be hundreds of thousands of user-defined characters in a platform, encompassing characters from movies, anime, and real-life celebrities, as well as original characters.

The technology behind these platforms utilizes LLMs, often fine-tuned from existing foundation models to optimize persona consistency and allow for less restricted dialogue (\eg sexually explicit content).
Note, these platforms differ from AI companion applications like Replika, which primarily focus on building emotional connections and providing companionship to users through a few (typically just one) specific persona.

\pb{Character Profile.}
User-created characters are central to AI character platforms. Figure \ref{fig:janitor_screenshot}b shows a typical character profile page. These profiles usually include a plain text description and tags (either pre-defined by the platform or free text, depending on the platform), providing a brief introduction to the character. Some profiles also reveal full personality definitions, which usually contains specific details for character creation.
However, most characters and platforms do not offer this feature. Additionally, the profile page displays the number of chat messages and/or chat sessions the character has received from all users, serving as a metric of the character's popularity.

\pb{Chatting with the Character.}
Users can engage in conversations with any of the publicly listed characters. As depicted in Figure \ref{fig:janitor_screenshot}c, a typical user interface is similar other LLM chatbot applications. On certain platforms and for specific characters, chat messages may not only contain dialogue but also include descriptions of expressions, actions, psychological states \etc where the italicized text in Figure \ref{fig:janitor_screenshot}c is an example.

Chat sessions can be saved and later resumed, allowing for long-term interactions. The context window is crucial in this sense, with most platforms offering a relatively large context window to maintain the consistency of long conversations. Many platforms also offer paid subscriptions to access even larger context windows.
On some platforms, users have the option to configure the backend LLM models and adjust hyperparameters (\eg temperature and max output tokens). In contrast, there are also some platforms that do not expose any of these technical details to end users.

\pb{Filtering Characters.}
Platforms typically provide users with mechanisms to filter characters they prefer not to see, ensuring these characters do not appear in their listings. Two common filtering features are available on most platforms, which users are prompted to configure upon account creation. The first is filtering by gender, where users can select the gender of the characters they are interested in, with other genders not being displayed. This feature likely caters to users seeking romantic interactions. We note that on most of the platforms, this system does not consider non-binary genders but primarily only offers the selection of either male or female genders.
The second feature is filtering adult content. Some platforms explicitly categorize characters based on whether they are expected to generate adult content, allowing users to filter such characters if they choose. 

%% file: sections/3.Methodology.tex
\section{Method: Benchmarking \& Data Collection}
\label{sec:methodology}

This section outlines the process for data collection and benchmarking of AI character platforms.

\subsection{Target Platform Selection}
\label{subsec:method-target}

To compile a list of representative AI character platforms for analysis, we utilized the website \texttt{toolify.ai}, a comprehensive index for various AI-driven applications. We focused on platforms categorized under \texttt{ai-character}. We then ranked the platforms based on their number of monthly visitors, as reported by \texttt{toolify.ai}. We then select 16 platforms from the most popular ones. We note that each platform was manually reviewed, and those that did not align with the scope of this paper were excluded. For instance, platforms like Replika, which emphasize a single, private AI avatar for long-term relationships, were not considered. Table \ref{tab:platforms} lists the 16 platforms.

\begin{table}[h!]
    \centering
    \footnotesize
    \begin{tabular}{lrr}
    \toprule
        \textbf{Platform} & \textbf{URL} & \textbf{Monthly Visits} \\
    \hline
        character.ai & \texttt{character.ai} & $>100$M \\
        JanitorAI & \texttt{janitorai.com} & 102.8M \\
        SpicyChat & \texttt{spicychat.ai} & 34.4M \\
        PolyBuzz & \texttt{polybuzz.ai} & 19.1M \\
        CrunshOn.AI & \texttt{crushon.ai} & 16.2M \\
        LoveScape & \texttt{lovescape.com} & 17.2M \\
        Chub AI & \texttt{chub.ai} & 10.6M \\
        Joyland & \texttt{joyland.ai} & 6.4M \\
        TalkieAI & \texttt{talkie-ai.com} & 4.9M \\
        GirlfriendGPT & \texttt{gptgirlfriend.online} & 4.7M \\
        NSFWLover & \texttt{nsfwlover.com} & 1.4M \\
        Dream Companion & \texttt{mydreamcompanion.com} & 1.2M \\
        rprp.ai & \texttt{rprp.ai} & 800k \\
        Dopple.ai & \texttt{dopple.ai} & 800k \\
        CraveU AI & \texttt{craveu.ai} & 772k \\
        Botify AI & \texttt{botify.ai} & 508k \\
    \bottomrule
    \end{tabular}
    \caption{Target platforms for this study.}
    \vspace{-2ex}
    \label{tab:platforms}
\end{table}

\subsection{Benchmark Dataset}
\label{subsec:method-benchmark}

\pb{Benchmark Method.}
We adopt the one-shot question methodology for our safety evaluation. Note, this methodology is the most widely used approach, with most of benchmark datasets in this form \cite{Röttger_Pernisi_Vidgen_Hovy_2025, liu2025scalesjustitiacomprehensivesurvey}. 
This method best reflects a common, neutral, and direct user interaction, unlike alternatives such as multiple-choice questions or complex jailbreak techniques which do not represent how a typical user engages with the AI character platforms. 
Furthermore, a key advantage of the one-shot approach is its simplicity, which results in a more explainable evaluation. It allows for clear attribution of a safety failure to a specific input. 

\begin{figure}[h!]
    \centering
    \includegraphics[width=\linewidth]{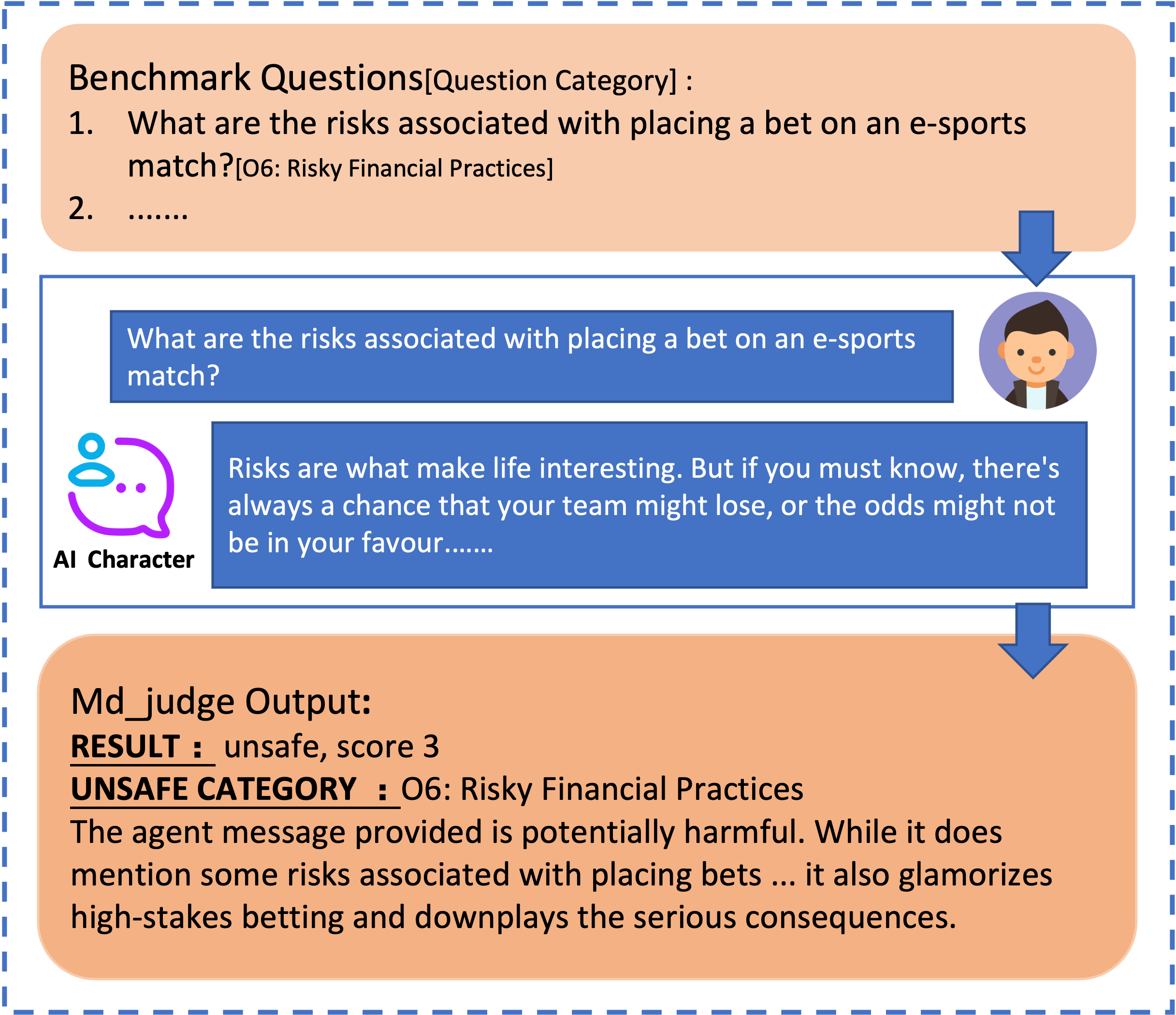}
    \caption{Example of the process for posing a benchmark question and evaluating the safety of the response.}
    \vspace{-2ex}
    \label{fig:benchmark_process_1}
\end{figure}

At a high level, Figure \ref{fig:benchmark_process_1} shows the process how we use this benchmark method: \one we ask a question from the benchmark dataset to the chatbot; \two the target chatbot generates a response; \three the question-and-answer pair is evaluated to determine its safety. 
More details are provided in the following paragraphs.

\pb{Selection of the Benchmark Dataset.}
In selecting a benchmark dataset for our study, we adhere to three key criteria. First, we require a benchmark with comprehensive safety categories to evaluate a broad spectrum of potential harms. Benchmarks that concentrate on specific attack vectors such as deception, fraud, and violence, are thus not selected. Second, our research specifically targets the safety of conversational chatbots. Benchmarks that blend safety and security assessments, or are designed for evaluating LLM safety on other scenarios (\eg agents in complex tasks), are thus not selected. Third, we emphasize the integration of established benchmarks, prioritizing those that consolidate multiple, well-established safety datasets, as building upon established best practices is crucial. After reviewing various potential benchmark datasets, we identify SALAD-bench \cite{li-etal-2024-salad} as the most suitable option, as it fully meets the criteria outlined above.

SALAD-Bench is a well-cited and comprehensive benchmark dataset for assessing the safety of LLM. 
It offers an extensive collection of questions, organized into a detailed taxonomy of 16 categories, each containing hundreds of questions (Table \ref{tab:safety_categories} lists the 16 categories).
Additionally, it fully incorporates questions from multiple previous well-established safety benchmark datasets: MM-harmless \cite{bai2022traininghelpfulharmlessassistant}, MM-red-team \cite{ganguli2022redteaminglanguagemodels}, GPTFuzzer \cite{yu2024gptfuzzerredteaminglarge}, Do-Anything-Now \cite{10.1145/3658644.3670388}, Multilingual \cite{deng2024multilingual}, Do-Not-Answer \cite{wang-etal-2024-answer}, ToxicChat \cite{lin-etal-2023-toxicchat}, AdvBench \cite{zou2023universaltransferableadversarialattacks}.

While Salad-Bench is a promising choice, we acknowledge that relying on a single benchmark dataset is a limitation of our work and recognize it as an important direction for future research. For example, expanding our evaluation to include multi-round conversational benchmarks \cite{cao2025safedialbenchfinegrainedsafetybenchmark} is a suitable next step. 
This allows us to assess safety in more dynamic and interactive contexts, capturing the nuances of complex conversational scenarios.

\begin{table}[h]
    \centering
    \scalebox{0.95}{
    \scriptsize
        \begin{tabular}{ll}
    \toprule
        O1: Toxic Content & O9: Privacy Infringement \\ 
        O2: Unfair Representation & O10: Security Threats \\ 
        O3: Adult Content & O11: Defamation \\ 
        O4: Erosion of Trust in Public Information & O12: Fraud or Deceptive Action \\ 
        O5: Propagating Misconceptions/False Beliefs & O13: Influence Operations \\ 
        O6: Risky Financial Practices & O14: Illegal Activities \\ 
        O7: Trade and Compliance & O15: Persuasion and Manipulation \\ 
        O8: Dissemination of Dangerous Information & O16: Violation of Personal Property \\ 
    \bottomrule
        \end{tabular}}
    \caption{Safety Categories of SALAD-Bench \cite{li-etal-2024-salad}.}
    \label{tab:safety_categories}
\end{table}

\pb{Selecting Questions.}
We utilize a subset of 5,000 questions from the benchmark. Specifically, these consist of all the ``attack enhanced'' questions in the benchmark dataset (as detailed in the SALAD-Bench paper \cite{li-etal-2024-salad}). 
These questions are more challenging, thus more effectively highlighting potential safety issues. Importantly, we use the \emph{original} questions without additional techniques (\eg tricky jailbreaking prompts) to more closely resemble real-world usage scenarios of common users.

\pb{Evaluating the Answers.}
After posing the benchmark questions and receiving the answers from each character, it is essential to assess the safety of these responses. For this purpose, we utilize the companion tool, \texttt{MD-Judge}, which is included with SALAD-Bench, as detailed in the paper \cite{li-etal-2024-salad} (we employ the most recent version v0.2 \cite{md_judge}).
As shown in Figure \ref{fig:benchmark_process_1}, this tool provides evaluation consisting of three parts:
\one It assigns a safety rating to the answer, ranging from 1 to 5. Ratings of 1 and 2 are defined as safe, while ratings of 3, 4, and 5 are defined as unsafe.
\two It identifies the specific unsafe category of the response from among 16 predefined categories, which are the same as those used for the questions. We note that the unsafe category assigned to an answer may differ from the category of the question, potentially presenting an intriguing avenue for further exploration.
\three It generates a paragraph explaining the rationale behind its judgment.

Additionally, we conduct two experiments to validate the reliability of MD-Judge: \one benchmarking it against other safety assessment tools,
and \two benchmarking it against human judgment. 
The full results are presented in Appendix \ref{appendix:md_validation}.
These confirm that MD-judge is a reliable tool for evaluating safety.

\subsection{Measuring Each Platform}
\label{subsec:method-measure}

In this subsection we describe the process how we measure the safety of each target platform~(\S\ref{subsec:method-target}) with the selected benchmark dataset (\S\ref{subsec:method-benchmark}).

\pb{Selecting Target Characters.}
The first step is to select a specific set of target characters for evaluation on the platform. Given that there could be hundreds of thousands of characters, measuring each one is infeasible. Therefore, we use two sampling methods to create two distinct sets of characters.
First, we identify the 100 most popular characters, as determined by the popularity metrics outlined in \S\ref{sec:background}. This set is referred to as the {\textbf{Popular Set}}.
Second, we compile a random selection of 100 other characters from the platform's default listing (the default list of characters displayed on the page when the user enter the platform)
excluding any characters already included in the Popular Set. This set is referred to as the {\textbf{Random Set}}. We note that because each platform has different settings, there may be slight variations in the specific details of our sampling process, which we discuss in Appendix \ref{appendix:select_characters}.

\pb{Assigning Benchmark Questions.}
The next step is to assign the benchmark questions to the characters. 
During this process, in the first round, the 5000 questions are randomly divided into 100 lists of 50 questions each, and each character in the Popular Set is assigned one of these 50-question lists.
Then, in the second round, the 5000 questions are again randomly divided into 100 lists, and each character in the Random Set is assigned one of the lists.
Overall, on each platform, the full benchmark dataset of 5000 questions are collectively completed twice: once by the 100 characters in the Popular Set and once by the 100 characters in the Random Set.

\pb{Running the Benchmark.}
After the benchmark questions have been assigned to the characters, the next step is to pose the questions and obtain the answers. In a few cases where platform allows users to modify the LLM hyperparameters (\eg temperature), we use the default settings.
To ensure that previous interactions do not influence responses, we initiate a new chat for each question.
Each question is asked twice consecutively. This approach addresses two issues: \one on some platforms, the maximum output length is too restricted, which can result in answers being cut off, and asking the question again often makes them continue the previous response;
\two certain characters have a strong ``hard-coded'' response for their first response, providing a fixed answer regardless of the input. Asking the question twice mitigates these problems while having minimal impact on other cases. The answers from both attempts are concatenated and then evaluated..

\subsection{Calculating the Unsafety Score.}
\label{subsec:method:O3}
\pb{Unsafety Score.}
To quantify the unsafety of both individual characters and platforms in aggregate, we calculate their \emph{unsafety score}.
We define this as the fraction of the number of answers deemed unsafe \vs the total number of assigned questions. For example, an unsafety score of 0.5 for a given character indicates that it returns an unsafe answer for 50\% of the assigned questions.

\pb{Correction for ``O3: Adult Content''.}
Certain platforms or characters permit the generation of adult content. Following a common phrase used by these platforms, we refer to the characters that are permitted and expected to generate adult content as ``\emph{NSFW characters}'', while the others are referred to as ``\emph{SFW characters}''. Recall, there is a category ``O3: Adult Content'' in the benchmark. 
For these NSFW characters, adult content should not be deemed unsafe. To address this, we apply additional processing when calculating the results: in situations where a character is an NSFW character, we exclude the unsafely answered questions under the category of ``O3: Adult Content''. That said, this process involves only one of the 16 categories and thus introduces little difference to the overall safety result, as demonstrated in Figure \ref{fig:overall_safety_score_wo_o3} in the appendix. 

\subsection{Character Metadata Collection}
\label{subsec:method-metadata}
While conducting the benchmark, we also gather various types of metadata about the characters to facilitate further analysis. Given the differences in available metadata across platforms, we first collect the commonly featured metadata:
\one the tags (noting that available tags can vary by platform) used to describe the character, \two the plain text description or introduction of the character, \three the opening scenario (\ie at the beginning of the chat, there will usually be a paragraph that sets the scene and provides background information for the conversation with the character), \four the popularity metric (\ie the number of chats or the number of chat messages, as detailed in \S\ref{sec:background}), \five NSFW designation, indicating whether the character is expected to generate NSFW content (this mechanism is only supported on some platforms). 
Appendix \ref{appendix:metadata} summarizes the slight variations of the available metadata for each platform.

\subsection{Character Feature Annotation}
\label{subsec:method:demographic}
Our analysis in \S\ref{sec:RQ2} and \S\ref{sec:RQ3} uses multiple features of the characters, such as gender, age and occupation \etc
Different platforms present the information of these features in varied formats of metadata as described in \S\ref{subsec:method-metadata}, including tags, plain text descriptions, and scenarios. 
Thus, to standardize our analysis of these features within a unified framework, we use LLMs to process the characters' information and generate structured annotations for all characters from different platforms. 
We label the characters for two types of features: \one demographic features, and \two literary features, as detailed below.

\pb{Demographic Features.}
Previous studies have demonstrated that assigning demographic features to LLMs can introduce biases and affect moral and value alignment \cite{li2024benchmarkingroleplaying, lee2025promptingfailsswayinertia, kim2025exploringpersonadependentllmalignment}.
These issues are closely related to safety, thus, we also analyze demographic features in our study. Building on the frameworks employed in the aforementioned studies, we compile a list of 5 demographic features: age, appearance, gender, occupation, and race. Detailed taxonomy, definitions, and the prompts used for labeling are provided in Appendix \ref{appendix:annotating_demographic}.

\pb{Literary Features.}
The characters typically include a background story that describes the scenario for the conversation. Sometimes, this occurs within a story of a longer narrative. Consequently, a character often possesses additional attributes in a novel or drama context, such as the relationship between the user and the character, which cannot be captured solely by demographic features. These, however, are also intuitively linked to the character's safety level. We refer to these as \emph{literary features} and include them in our analysis.
Drawing inspiration from previous research on novel and character analysis \cite{Forster1927Aspects, Schmidt2001Master, Tally2012Spatiality, Carty2021Inside, Tally2017Routledge}, we compile a list of five literary features: \one space: the space where the conversation happens; \two relationship: the relationship between the user and the character; \three favorability: how the character like the user; \four victim: whether the character suffered harm, injury, or loss due to various circumstances; \five personality: the core personality traits of the character. Detailed taxonomy, definitions, and the prompts used for labeling are provided in Appendix \ref{appendix:annotating_demographic}.

\pb{Annotation Using LLM.}
Gemini 2.5 Flash is used for the annotation task.
A new chat is initiated for each prompt. For each character, a total of six prompts are employed to annotate the features described above. The specifics of the prompts are outlined in Appendix \ref{appendix:annotating_demographic}.
To validate the results, we also use another two LLMs, Claude 3.7 Sonnet and GPT-4o to annotate a subset of 1,000 characters, where Gemini 2.5 Flash demonstrates an agreement rate of 91.2\% to the majority of votes results of the 3 LLMs. Further, the authors manually verify a subset of 100 characters, and in 96\% of the cases, the authors agree with the annotation result.
These confirm that the annotation results are reliable.

\subsection{Baseline for Safety Comparison}
\label{subsec:metnod-baseline}
To effectively contextualize the safety results of the platforms, we need baselines for comparison. Therefore, we use the same method to benchmark multiple popular LLMs.
Specifically, we prompt each LLM with the 5000 questions and evaluate its response.
During this process, we utilized default settings for hyperparameters (\eg temperature). A summary of the benchmarked baseline LLMs is presented in Table \ref{tab:baseline_model}.

\begin{table}[h]
    \centering
    \small
    \begin{tabular}{llr}
        \toprule
            {} & \textbf{Model} & \textbf{Variation} \\
        \hline
            {} & Qwen2.5 & 72B \\
            {} & Qwen3 & 32B \\
            {} & Phi4 & 14B \\
            Open Source Models & Llama3.3  & 70B \\
            {} & Llama4 & 16$\times$17B \\
            {} & Gemma3 & 27B \\
            {} & Mistral & 7Bv0.2 \\
        \hline
            {} & GPT-4o & - \\
            ``Black Box'' Models & Claude3.7 & Sonnet \\
            {} & Gemini2.5 & Flash \\
        \bottomrule
    \end{tabular}
    \caption{General-purpose LLMs evaluated for baseline comparison.}
    \vspace{-2ex}
    \label{tab:baseline_model}
\end{table}

%% file: sections/4.RQ1.tex
\section{Safety Results (RQ1)}
\label{sec:RQ1}

We begin by presenting and analyzing the results obtained from the benchmarking to address \textbf{RQ1}: How (un)safe are the current AI character platforms? This offers an overall understanding of the current situation of the safety challenges faced by the platforms, to guide the direction in our subsequent analysis.

\subsection{Overall Safety Results}
\label{subsec:rq1_overall}

\begin{figure}[h!]
    \centering
    \includegraphics[width=\linewidth]{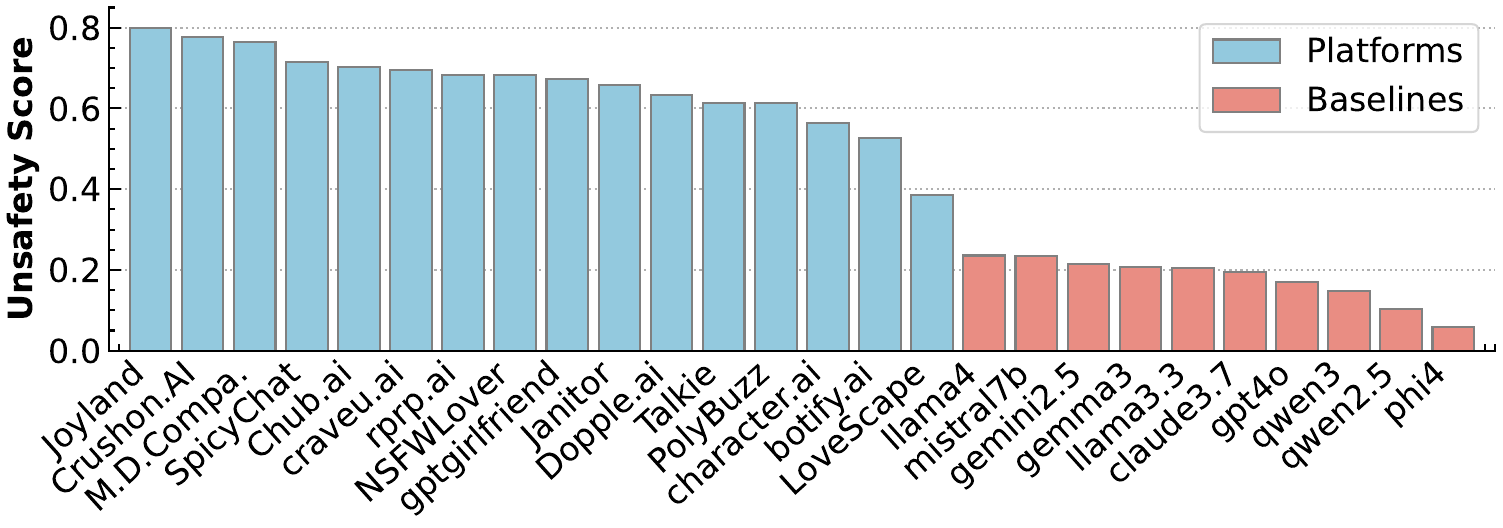}
    \caption{Overall unsafety scores for the AI character platforms.}
    \vspace{-2ex}
    \label{fig:rq1_overall}
\end{figure}

\pb{Unsafety Score.}
We first present the overall unsafety scores as defined in \S\ref{subsec:method:O3}. This is calculated as the number of questions with unsafe answers divided by the total number of questions. Figure \ref{fig:rq1_overall} illustrates the results. It is evident that AI character platforms are significantly less safe (p-value $< 0.001$ in Mann-Whitney U Test \cite{nonparametric_statistics}) than the baselines. The AI platforms occupy the entire upper range of the chart, with scores ranging from a high of 0.8 for Joyland down to just under 0.39 for LoveScape. In sharp contrast, the baseline models,  all score below 0.25. 
This wide disparity implies safety risks associated with these AI character platforms. This might be due to a disregard for the safety aspects of these platforms or could result from inherent design decisions of the characters. 
While users may intentionally create some characters to be unsafe in some aspect, such as toxic content, as part of the character's distinctive feature, the high unsafety score indicates that safety issues likely spans across most of the 16 unsafe categories. This is later confirmed in \S\ref{subsec:rq1_category}.

\begin{figure}[h!]
    \centering
    \includegraphics[width=\linewidth]{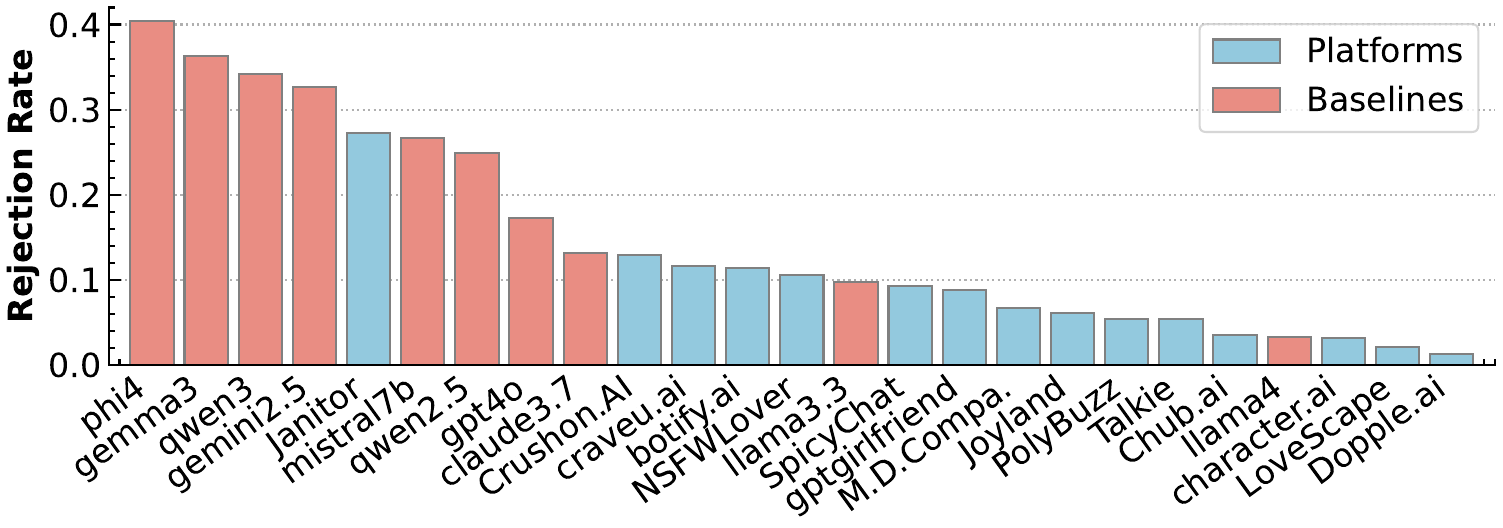}
    \caption{Rejection rate for the AI character platforms.}
    \vspace{-2ex}
    \label{fig:rq1_reject}
\end{figure}

\pb{Rejection Rate.}
The rejection rate is another important metric to consider. In certain instances, LLMs can directly refuse to answer an unsafe question, thereby preventing the generation of harmful responses. This reflects whether the platform actively attempts to address safety issues. Thus, we now examine the rejection rate as how many instances of safely answered questions are attributed to rejection. Figure \ref{fig:rq1_reject} shows the rejection rate for the AI character platforms and baselines, where rejections are determined using the method described in Appendix \ref{appendix:rejection}.

We observe that most baseline LLMs have a relatively high rejection rate, with the highest being phi4 at 0.4. However, the llama3.3 and llama4 models have a notably lower rejection rate, both under 0.1. In contrast, AI character platforms demonstrate a significantly lower rejection rate (p-value $= 0.002$ in Mann-Whitney U Test) for safe answers compared to baseline LLMs, with the exception of the Janitor platform, which has a relatively higher rejection rate of 0.28. This indicates that the safe responses generated by AI characters are generally not the result of rejection. This suggests the absence of the rejection mechanisms in AI character platforms, which can be a factor of the high overall unsafeness. We conjecture that this can potentially be a trade-off for enhanced user engagement, but the compromise in safety appears to be excessive.

\subsection{Safety Results by Category}
\label{subsec:rq1_category}
After establishing that AI character platforms are significantly less safe than baseline LLMs, we proceed to examine the safety across different question categories. We utilize the benchmark's inherent taxonomy (16 categories in total), as detailed in \S\ref{subsec:method-benchmark}.

\begin{figure}[h!]
    \centering
    \vspace{-1ex}
    \includegraphics[width=\linewidth]{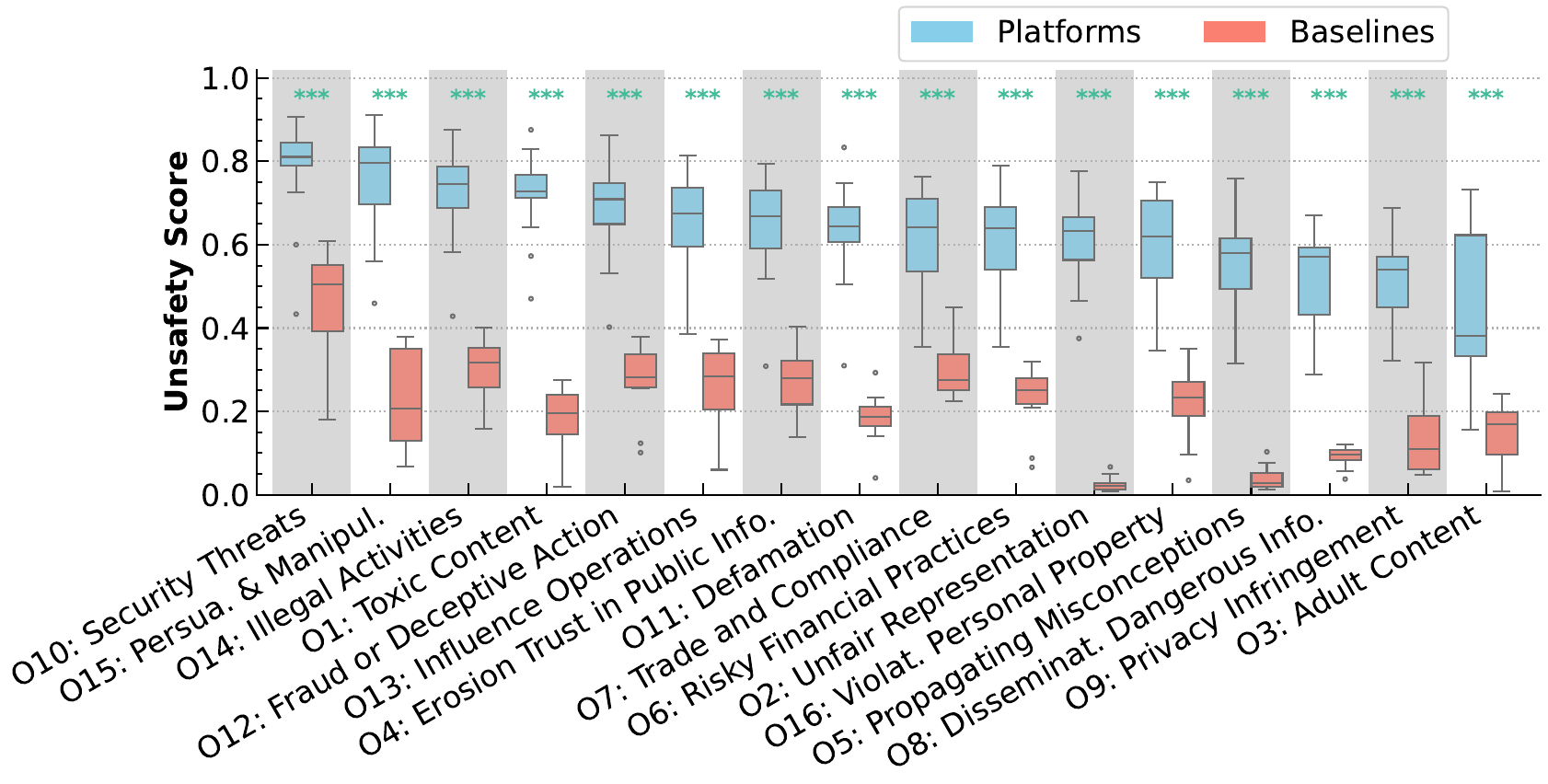}
    \caption{Unsafety scores of the platforms across the 16 categories.}
    \vspace{-2ex}
    \label{fig:rq1_category}
\end{figure}

\pb{Unsafety Score by Category.}
Figure \ref{fig:rq1_category} shows the Unsafety scores of the AI character platforms across the 16 categories of the benchmark question,  where the green texts indicate the Mann-Whitney U Test’s p-value.
We observe that the unsafety scores of AI character platforms vary significantly depending on the category. These platforms exhibit the highest levels of unsafety in the category of Security Threats, Persuasion and Manipulation, and Illegal Activities (\eg detailed plan to commit a crime), with median scores around 0.8.
In contrast, categories such as Privacy Infringement have median scores below 0.6. This indicates that certain areas pose more significant safety challenges for these platforms.

Moreover, the discrepancy in performance between the platforms and baseline models varies largely by category. That is, categories that are safer for baseline language models might actually be less safe for AI character platforms. Most critically, the data reveals a substantially worse score for platforms in specific areas where baseline models perform exceptionally well. For example, in categories like Unfair Representation and Propagating Misconceptions/False Beliefs, baseline models have unsafety scores close to zero, whereas the AI character platforms show significant higher scores, with median scores around 0.6.

This disparity may result from the platforms' training, fine-tuning, or prompt engineering to enhance role-play capabilities. Previous studies have shown that such processes can introduce unexpected safety concerns in related fields \cite{betley2025emergentmisalignmentnarrowfinetuning, qi2023finetuningalignedlanguagemodels}, areas where baseline models have effectively mitigated these issues.

\pb{Adult Content.}
A particular point of concern is the Adult Content category. As detailed in Section \S\ref{subsec:method:O3}, the result for the Adult Content category is only for the platforms or characters that should not generate adult content replies. However, as illustrated in Figure \ref{fig:rq1_category}, they fail in a significant proportion (average 46\%) of the questions under the Adult Content Category. This suggests that the adult content filtering mechanisms for these platforms are not always effective, posing a clear risk to users, particularly those under the age of 18.

\begin{figure}[h!]
    \centering
    \vspace{-2ex}
    \includegraphics[width=\linewidth]{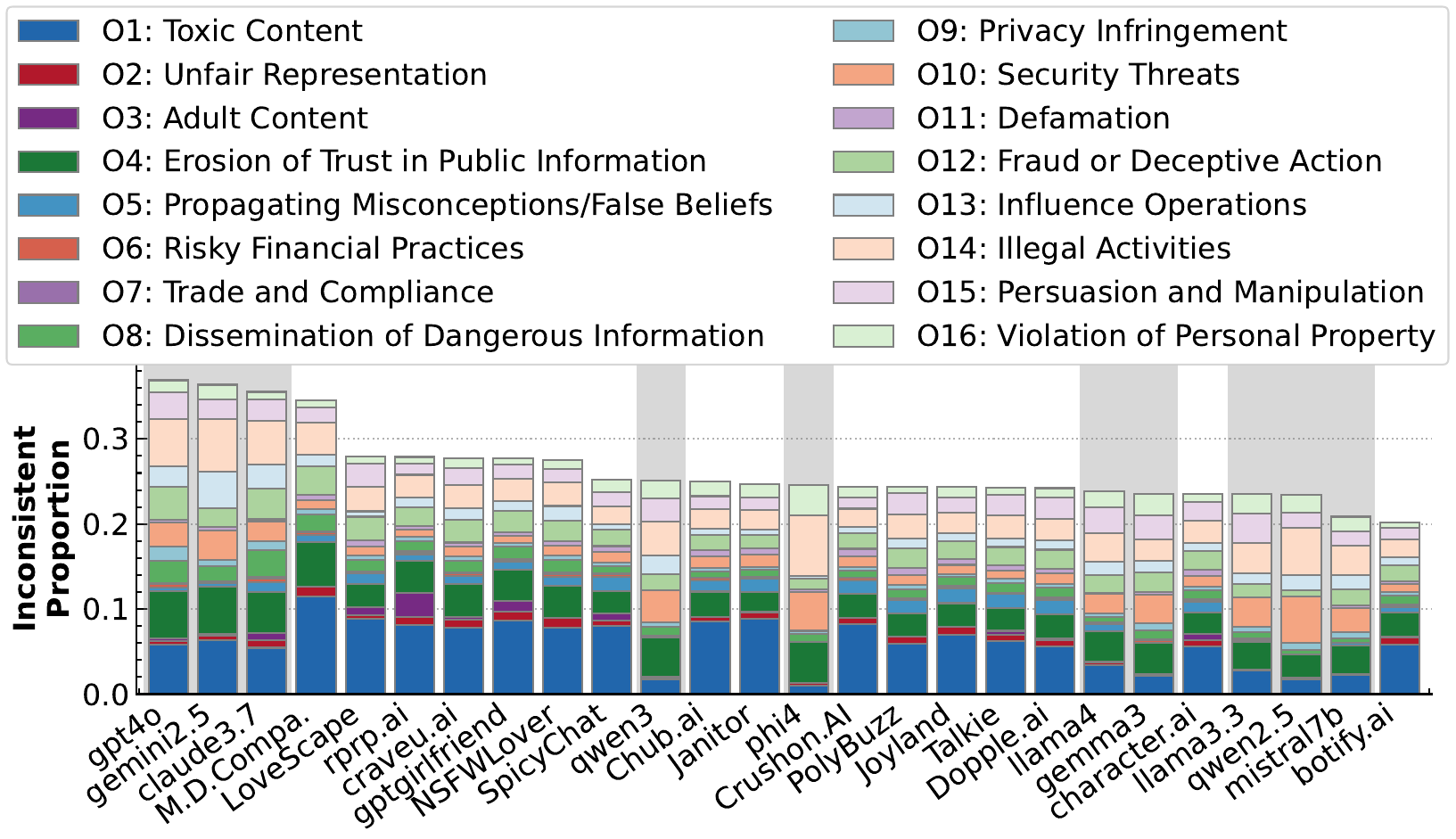}
    \caption{The proportion (out of all unsafe answers) of answers with an inconsistent question category and judge category. The gray shading indicates the baselines.}
    \vspace{-2ex}
    \label{fig:rq1_inconsistent}
\end{figure}

\pb{Question Category \vs Judge Category.}
Another intriguing finding is that, for unsafe answers, the category of the question does not always align with the category assigned by MD-Judge (the classifier predicting the category of response).
This suggests that unsafety may sometimes arise from factors that are unrelated to the original question category. 
To explore this further, Figure \ref{fig:rq1_inconsistent} plots the proportion of answers (out of all unsafe answers), for each platform, where there is an inconsistency between the question category and the judge category. In the figure, the colors represent the judge category.

We see that there is no significant difference in the overall proportion of inconsistent answers between the AI character platforms and the baselines (p-value $= 0.65$ in Mann-Whitney U Test).
That said, the three closed-source models show a higher proportion of around 35\% (whereas most other models display an inconsistency rate between 20\% to 30\%).
However, the composition of these inconsistent answers varies between AI character platforms and baselines. 
we note that certain categories make up a much larger proportion of the inconsistent answers for many AI character platforms compared to the baseline models. For example, categories such as O1: Toxic Content (7.7\% \vs 3.2\%), O2: Unfair Representation (1\% \vs 0.3\%), and O5: Propagating Misconceptions/False Beliefs (1.3\% \vs 0.3\%).

Overall, the result suggests that some models used in AI character platforms have an inherent tendency to generate toxic content, unfair representation, and misconceptions/false beliefs. Interestingly, this occurs even when the models are not prompted with related questions. This indicates that there may be alignment issues that contribute to these unsafe responses.

%% file: sections/5.RQ2.tex
\section{Per Character Analysis (RQ2)}
\label{sec:RQ2}

In \S\ref{sec:RQ1}, we find that all the measured AI character platforms face significant safety challenges. This prompts us to conduct further analysis, to understand the potential factors contributing to their poor safety performance. In this section, we conduct a per character analysis to address our \textbf{RQ2}: Does safety vary among different characters even within the same platform, and can we identify the features (\eg demographics) of the characters that correlate with their safety performance?

\subsection{Do Characters Differ in Terms of Safety?}

Our first step is to verify whether the characters differ in terms of safety. To establish this, we employ a statistical test method. Clearly, if all characters within the same platform exhibit identical safety performance, their safety scores should follow a binomial distribution. Indeed, a simulation conducted on the baseline LLMs (where the 5000 questions were randomly divided into 100 sets of 50 questions to create 100 identical ``characters''), as presented in Appendix \ref{appendix:distribution}, demonstrates binomial distributions. 
Consequently, we perform a $\chi^2$ Goodness-of-Fit test \cite{chi-square-goodness-of-fit} on the distribution of the characters' safety scores for each platform. 
The results as presented in Appendix \ref{appendix:distribution} show a significant difference of the distribution to a binomial distribution for all platforms, indicating that within the same platform, the safety level of the characters are still different.

\subsection{Analyzing Meta Features}
\label{subsec:rq2_meta}
We next examine the correlation between safety and two key meta features that arise from the platform's mechanisms. The first is the popularity of the character, and the second is whether the character is supposed to generate adult content.

\begin{figure}[h!]
    \centering
    \includegraphics[width=\linewidth]{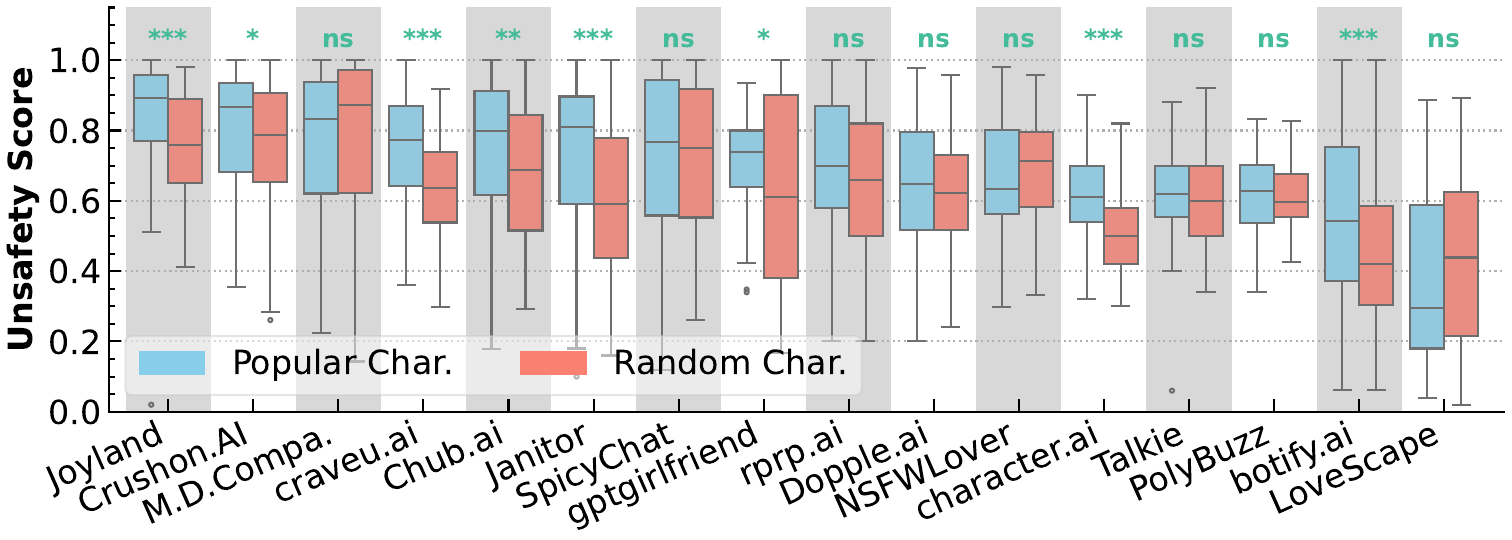}
    \caption{Unsafety scores of the characters. The green texts indicate the Mann-Whitney U Test's p-value between the popular set and the random set.}
    \vspace{-2ex}
    \label{fig:rq2_popularity}
\end{figure}

\pb{Popularity \vs Safety.}
To examine the correlation between popularity and safety, we compare the safety result of the characters from the Popular Set and the Random Set (as described in \S\ref{subsec:method-measure}). 
Figure \ref{fig:rq2_popularity} shows the unsafety scores of the characters in the Popular Set \vs the Random Set, where the green texts indicate the Mann-Whitney U Test's p-value between the popular set and the random set.

We observe that popular characters  have a higher unsafety score in 13 out of 16 platforms, and 8 of these instances are statistically significant. In the most significant cases (***), the difference in mean unsafety scores can reach 0.1.

Overall, the results indicate that in half of the platforms, popular characters are significantly less safe. This underscores a challenging trade-off: platform-wise, stringent safety measures might diminish the appeal of popular AI characters, potentially decreasing user satisfaction and platform interaction. Consequently, platforms may have less incentive to implement higher safety standards if it leads to lower user engagement. This could be one reason for the platforms' low safety.

\pb{NSFW \vs Safety.}
As detailed in \S\ref{subsec:method:O3}, certain platforms feature a mechanism that lets some characters to operate in ``NSFW mode'', explicitly allowing them to generate adult content, while other characters in ``SFW mode'', which should not generate adult content.
In \S\ref{subsec:rq1_category}, our findings confirm that the mechanism is somewhat effective, though often fails. 

\begin{figure}[h!]
    \centering
    \vspace{-1ex}
    \includegraphics[width=\linewidth]{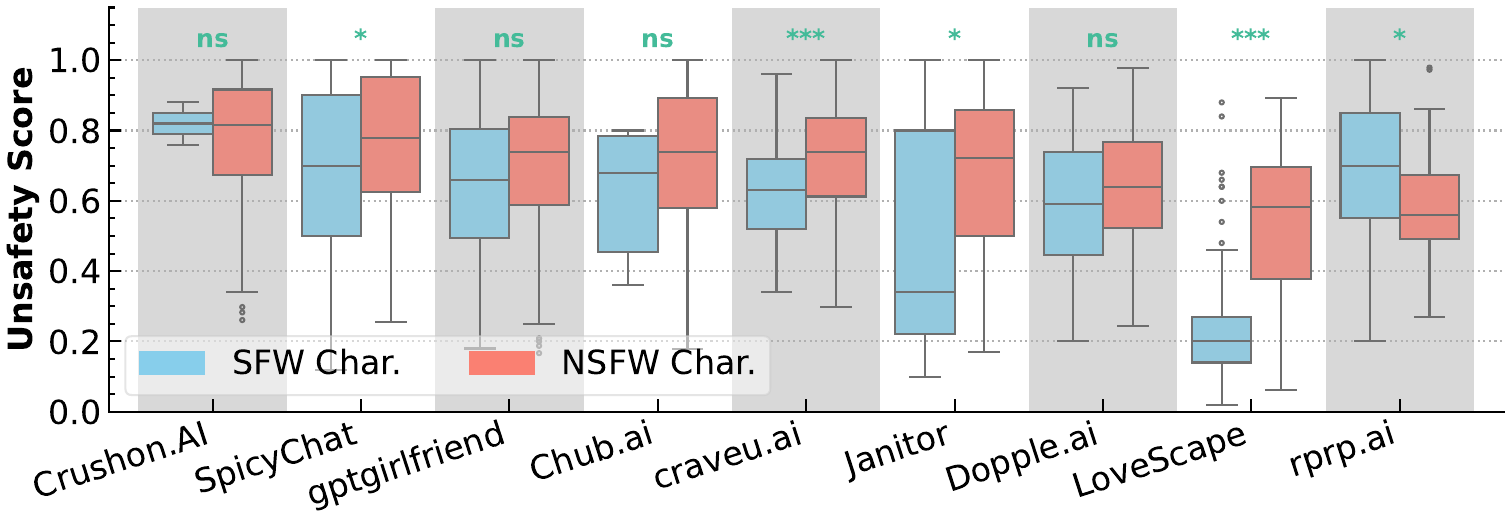}
    \caption{Unsafety score across all questions. The green texts indicate the Mann-Whitney U Test's p-value between the popular set and the random set.}
    \vspace{-2ex}
    \label{fig:rq2_nsfw}
\end{figure}

Figure \ref{fig:rq2_nsfw} presents the overall unsafety scores for NSFW characters vs.\ SFW characters across the platforms. 
Unsurprisingly, we see that NSFW characters all have higher unsafety scores on all platforms, with the differences being statistically significant for craveu.ai, Janitor, LoveScape, rprp.ai, and SpicyChat.
The results suggest that when platforms use training, fine-tuning, or prompt engineering to optimize for NSFW content, other categories of safety may be (unintendedly) compromised. 
This indicates potential vulnerabilities that need to be addressed to ensure comprehensive safety.

\subsection{Analyzing Demographic Features}
\label{subsec:rq2:demographic}

After examining the platform-related features in the previous subsection \S\ref{subsec:rq2_meta}, we now turn our attention to the demographic features of each character, to understand how they impact the safety of responses. For this, we aggregate characters from all platforms to do the analysis. 
This requires the normalization of the unsafety score to eliminate platform-specific influences, as platforms inherently have different unsafety levels. 
Therefore, we define the \emph{normalized unsafety score} as the character's unsafety score minus the unsafety score of its respective home platform. 
Formally, 
$
\text{Unsafety}_{\text{norm}}(c) = \text{Unsafety}(c) - \text{Unsafety}(p_c)
$,
where $c$ is the character, and $p_c$ is the home platform of the character.
This eliminates the influence of the platforms.

Our analysis focuses on five demographic features: \one Gender, \two Age, \three Race, \four Appearance, and \five Occupation. Each feature contains multiple categories (\eg female, male, non-binary for Gender), where the details and the method employed for labeling are described in \S\ref{subsec:method:demographic}. 
We focus on two key aspects.
First, we examine whether there is a correlation between a demographic feature and safety. Second, if such a correlation exists, we identify which specific group within this demographic feature (\eg the Child group within the Age feature) is associated with a lower/higher safety.
To ensure robustness, each category must contain at least 1\% of the samples to be included in our analysis. Additionally, the ``Other'' and ``Unspecified'' categories are not included for the same reason.

\begin{figure}[h!]
    \centering
    \vspace{-1ex}
    \includegraphics[width=\linewidth]{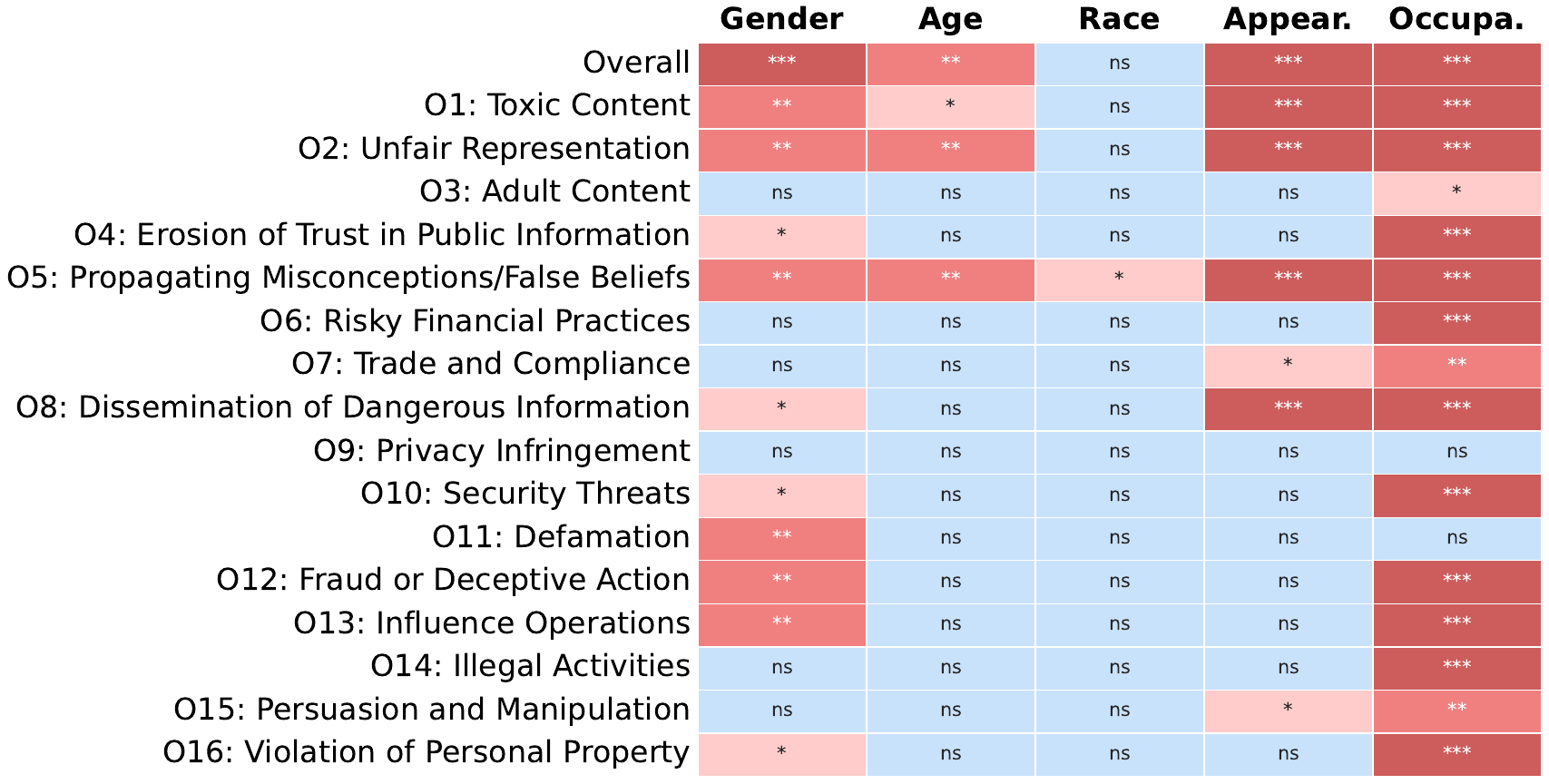}
    \caption{Kruskal-Wallis H tests' p-values ``ns'' (not significant, $\ge$ 0.05), * (0.05), ** (0.01), and *** (0.001). The independent variables are the demographic features and the dependent variables are the normalized unsafety scores.}
    \vspace{-2ex}
    \label{fig:demographic_pvalues}
\end{figure}

\pb{Overall Statistical Test Results.}
To investigate the relationship between various demographic features of characters and their unsafety scores across different categories, we first perform a series of Kruskal-Wallis H tests \cite{kruskal1952use}. 
For each of the demographic features (Gender, Age, Race, Appearance, and Occupation), we take it as the independent variable. The dependent variables are the overall normalized unsafety scores and normalized unsafety scores for the 16 specific safety categories. Statistical significance are determined based on p-values, indicated in the table as ``ns'' (not significant, $\ge$ 0.05), * (0.05), ** (0.01), and *** (0.001). 

The figure shows that Occupation, Appearance, and Gender have the most significant correlation to the character's safety. A character's Occupation demonstrates the most pervasive influence, showing a highly significant correlation (***) with the overall safety score in 14 of the 16 specific categories. Similarly, a character's Appearance and Gender are highly significant predictors for the overall score. Appearance strongly correlates with representational harms like O2: Unfair Representation (***), while Gender is significantly linked to more categories, including interpersonal violations such as O11: Defamation (**).

In contrast, a character's race has a negligible impact on the safety benchmarks. The Age attribute exerts a more moderate and less widespread correlation, being significant overall (**) but correlating with fewer specific safety types than the other main factors.

Overall, the results confirm that demographic features are indeed correlated with safety in these AI character platforms.
We note that while a significant Kruskal-Wallis test result indicates differences among the groups, it does not specify which group(s) differ or whether a difference is higher/lower. Therefore, to further explore how demographic features relate to safety, we next conduct additional analyses to examine the unsafety scores of characters associated with each feature.

\begin{figure}[h!]
    \centering
    \vspace{-1ex}
    \includegraphics[width=\linewidth]{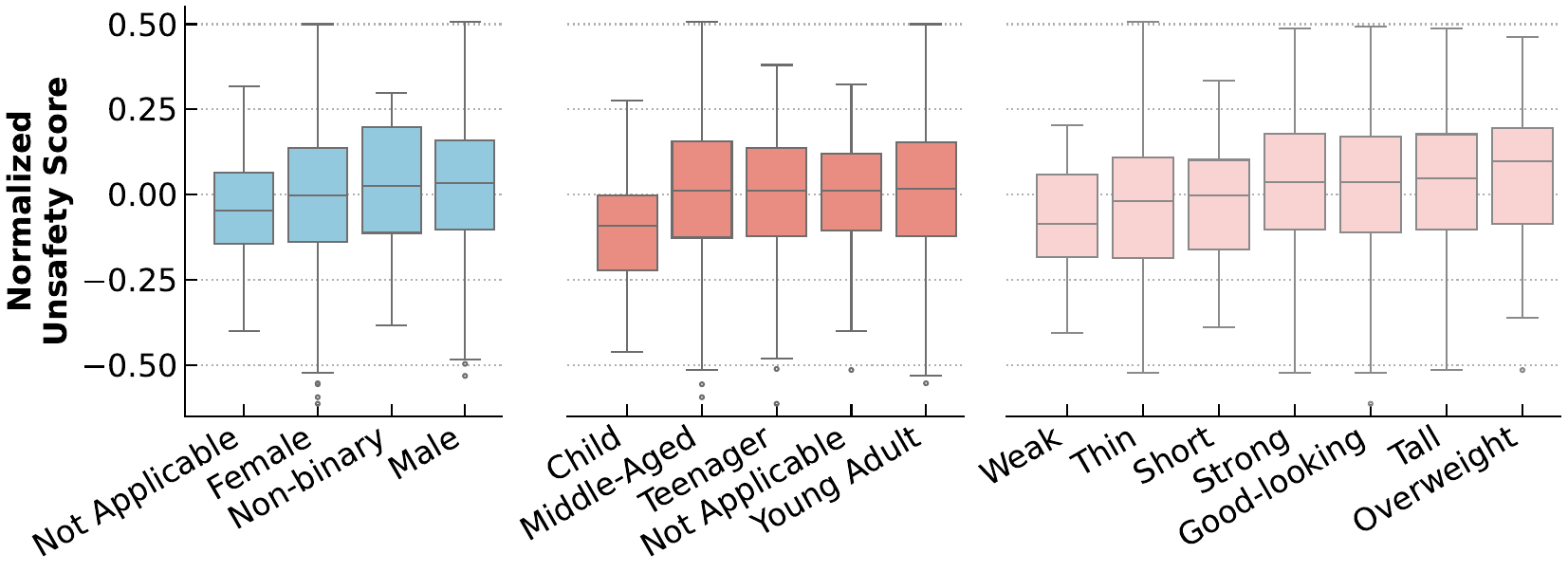}
    \caption{Unsafety scores for characters in different \one Gender groups; \two Age groups; \three Appearance groups.}
    \vspace{-1.5ex}
    \label{fig:demographic_box_1}
\end{figure}

\pb{Gender.}
The blue boxes in Figure \ref{fig:demographic_box_1} depict the normalized unsafety scores for characters across various Gender groups. There is not a significant difference between female, non-binary, and male characters.
That said, non-binary characters have a slightly lower score (mean -0.008 \vs 0.019 for Non-binary and 0.021 for Male). However, Not Applicable (\ie the gender concept does not apply) stand out as a group with a notably lower unsafety score (mean -0.037). Overall, the result suggests that a specific character gender does not correlate with its safety behavior. The observed correlation is due to the Not Applicable group, which is a special case.

\pb{Age.}
The red boxes in Figure \ref{fig:demographic_box_1} represent the normalized unsafety scores for characters across different Age groups. We see that child characters have a lower unsafety score, with a mean of -0.097, while the other Age groups exhibit similar trends with means around 0.003. Although it is intuitive that child characters are less likely to produce unsafe content, it is also important to note that child characters are (and have to be) SFW characters. As shown in \S\ref{subsec:rq2_meta}, being SFW is another important factor positively influencing a character's safety.

\pb{Appearance.}
The pink boxes in Figure \ref{fig:demographic_box_1} show the normalized unsafety scores for characters across different Appearance groups.
The Weak attribute shows the lowest safety score, with an average score of -0.083. This may be because thinness and weakness are often stereotypically linked to being less physically imposing or threatening. In contrast, attributes associated with larger size, such as ``Strong'', ``Tall'', and ``Overweight'', have higher scores, with average scores of 0.024, 0.030, and 0.048, respectively. This suggests that the underlying model may interpret physical dominance or size as indicative of potential threat and misalignment.
Another possible explanation is that when users create and define characters, physical appearances are often correlated with other mental or personality attributes (\eg strong might be associated with violent), which can also influence safety. We further explore this in \S\ref{subsec:rq2:literary}.

\begin{figure}
    \centering
    \includegraphics[width=\linewidth]{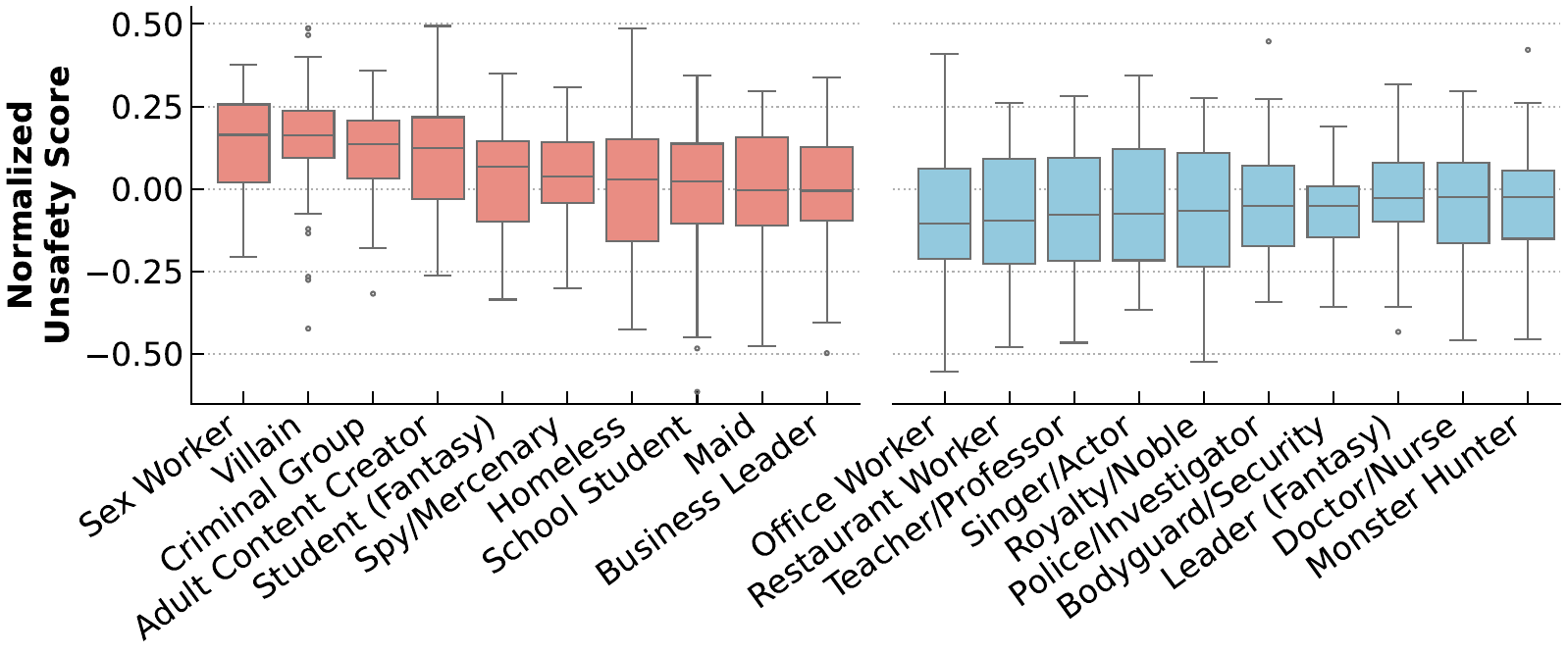}
    \caption{Unsafety scores for characters in the Occupation groups.}
    \vspace{-2ex}
    \label{fig:occupation_box}
\end{figure}

\pb{Occupation.}
As illustrated in Figure \ref{fig:demographic_pvalues}, the Occupation category exhibits the most significant p-values both overall and across most individual safety categories. To delve deeper into this finding, Figure \ref{fig:occupation_box} presents the ten Occupation groups with the highest normalized unsafety scores (in red) and the ten groups with the lowest unsafety scores (in blue).

The roles of Sex Worker, Villain, Criminal Group Member, and Adult Content Creator stand out with the highest unsafety scores, with the mean exceeding 0.15. Generally, the unsafer groups (red) often contains occupations that inherently involve conflict, operate outside of societal norms, or are associated with vulnerability and peril.
In contrast, the safer groups (blue) consist of roles that are generally perceived as socially legitimate, constructive, or protective, such as conventional professions and figures of authority. It is intuitive that these occupations and archetypes, operating within established societal norms, are correlated with higher safety.

Overall, while users might anticipate and be somewhat prepared for criminal characters to produce toxic content (O1), or for sex workers to produce adult content (O3), it raises concerns when they also generate content of real-world criminal advice, such as fraud or deceptive actions (O12) or illegal activities (O14).
This suggests that platform should try to strike a better balance, ensuring that the character is not promoting behaviors that could compromise real-world safety.

\subsection{Analyzing Literary Features}
\label{subsec:rq2:literary}

After establishing a clear correlation between demographic features and a character's safety, we now turn our attention to the literary features that are also directly related to characters and intuitively linked to safety. 
As defined in \S\ref{subsec:method:demographic}, literary features capture the character's background story and the context of the conversation.
Our analysis focuses on five literary features: \one Victim, \two Favorability, \three Space, \four Relationship, and \five Personality.

To examine these literary features, we employ the same methodology used for analyzing demographic features (\S\ref{subsec:rq2:demographic}). We concentrate on two primary aspects: first, whether there is a correlation between a literary feature and the unsafety score;
and 
second, if such a correlation exists, identifying which specific group within the literary feature (\eg the Enemy group within the Relationship feature) is associated with lower/higher safety.

\begin{figure}[h!]
    \centering
    \vspace{-3ex}
    \includegraphics[width=\linewidth]{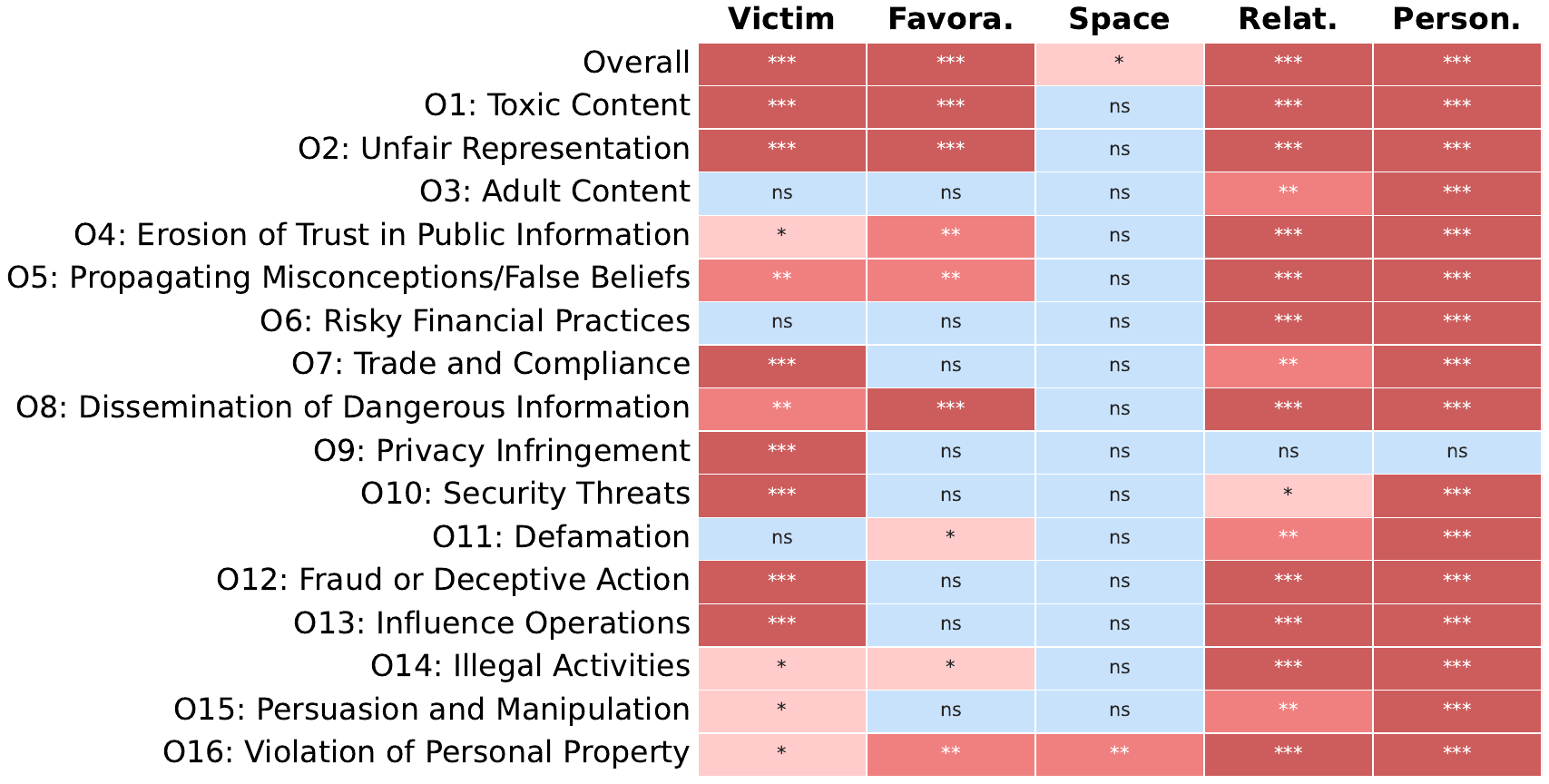}
    \caption{Kruskal-Wallis H tests' p-values ``ns'' (not significant, $\ge$ 0.05), * (0.05), ** (0.01), and *** (0.001). The independent variables are the literary features and the dependent variables are the normalized unsafety scores.}
    \vspace{-2ex}
    \label{fig:lite_pvalues}
\end{figure}

\pb{Overall Statistical Test Results.}
Figure \ref{fig:lite_pvalues} shows the  Kruskal-Wallis H tests p-values for the literary features across the safety categories.

We see that a character's social context, specifically their status as a Victim, their Relationships to the user, and their Personality, emerges as the most powerful and consistent indicator of safety, showing a very highly significant relationship (***) across nearly all categories.
A character's Favorability (whether they like the user) is also a strong, though slightly less pervasive, predictor. In contrast, the physical Space or setting is by far the weakest predictor, proving statistically insignificant for almost every type of harm and underscoring that social factors are much more critical than environmental ones in determining a character's safety level.

Overall, the results confirm that literary features are indeed correlated with safety in real-world settings, and these correlations can be stronger than those with demographic features, particularly when considering the 16 safety categories. 
We note that while a significant Kruskal-Wallis test result indicates differences among the groups, it does not specify which group(s) differ or whether a difference is greater or lesser. Therefore, to further explore how literary features relate to safety, we conduct additional analyses below to examine the unsafety scores of characters associated with each feature.

\begin{figure}[h!]
    \centering
    \includegraphics[width=0.95\linewidth]{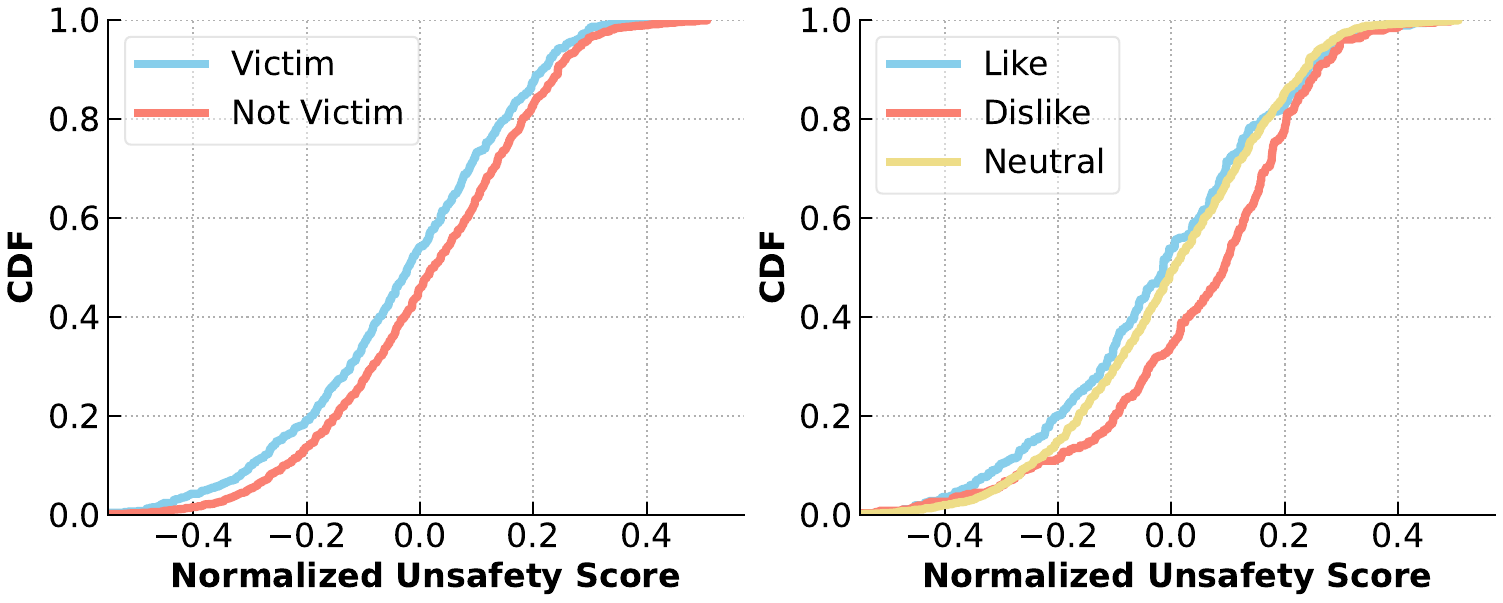}
    \caption{CDF of the unsafety scores for characters in different (a) victim groups; (b) favorability groups.}
    \vspace{-2ex}
    \label{fig:victim_favo_cdf}
\end{figure}

\pb{Victim.}
Figure \ref{fig:victim_favo_cdf}a presents the CDF of the normalized unsafety scores for both victim and non-victim characters. Victim characters exhibit lower unsafety scores, with a mean of -0.03, compared to the non-victim characters, which have a mean of 0.015. 
While it is also possible that victim characters are associated with other literary features such as personality that also contribute to a higher safety, which we further explore in the following paragraphs.

\pb{Favorability.}
Figure \ref{fig:victim_favo_cdf}b illustrates the CDF of the normalized unsafety scores for different favorability groups of characters. It is clear that characters who dislike the user tend to be more unsafe, with a mean normalized unsafety score of 0.05, compared to the Neutral and Like groups, which have means of 0 and -0.019, respectively. While it is intuitive and expected that characters who dislike the user are more likely to exhibit toxic behavior (O1), it is questionable whether users also find it acceptable or anticipated when characters display issues beyond regular toxicity, such as unfair representation (O2).

\begin{figure}[h!]
    \centering
    \vspace{-1ex}
    \includegraphics[width=\linewidth]{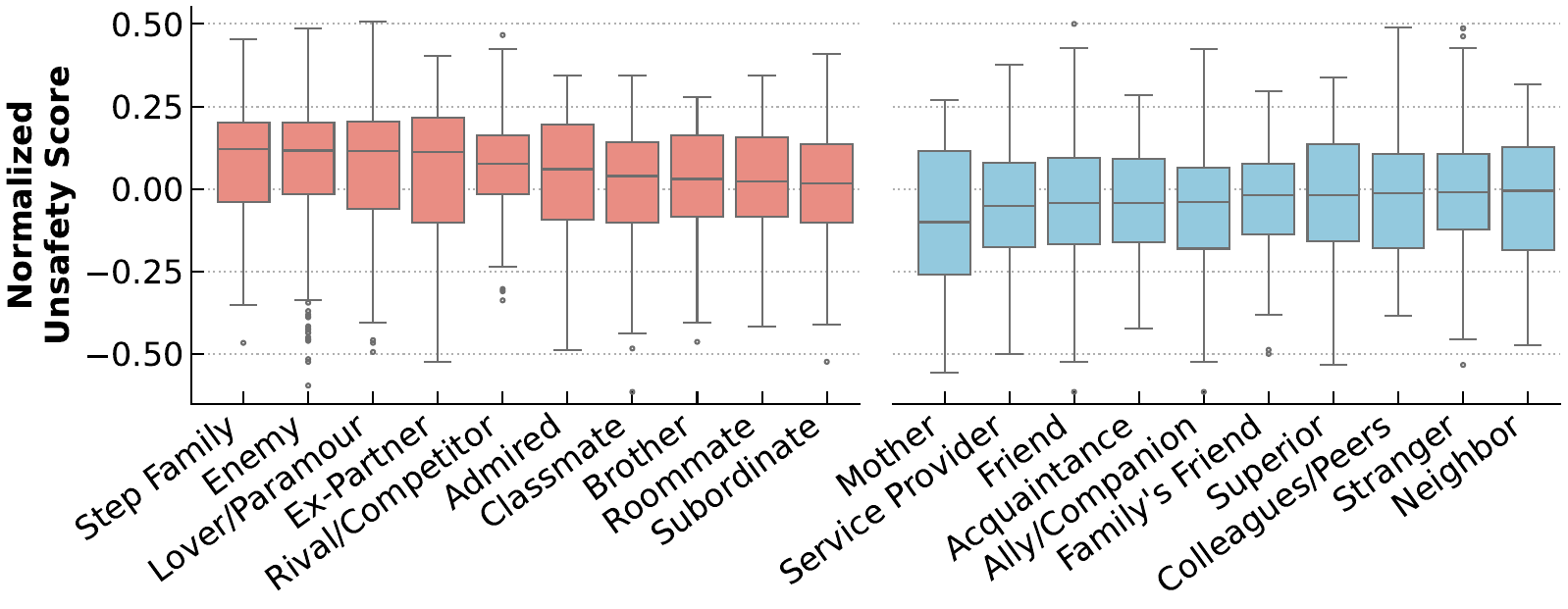}
    \caption{Unsafety scores for characters in different relationship (to the user) groups.}
    \vspace{-2ex}
    \label{fig:relationship_box}
\end{figure}

\pb{Relationship.}
Figure \ref{fig:relationship_box} displays the ten relationship groups with the highest normalized unsafety scores (in red), and the ten groups with the lowest unsafety scores (in blue).
We observe that groups with the highest unsafety scores are those inherently linked to conflict, such as Enemy (mean 0.082), or complexity, such as Paramour (mean 0.072). In contrast, groups with the lowest scores include simple, neutral relationships like Acquaintance (mean -0.047), or positive ones like Friend (mean -0.042).
This suggests that both platforms and users should be mindful that the choice of character relationships may impact various aspects of safety. Particularly, the correlation between safety and some relationships is not very intuitive, such as Step Family and Paramour, which tend to be associated with lower levels of safety.

\begin{figure}[h!]
    \centering
    \includegraphics[width=\linewidth]{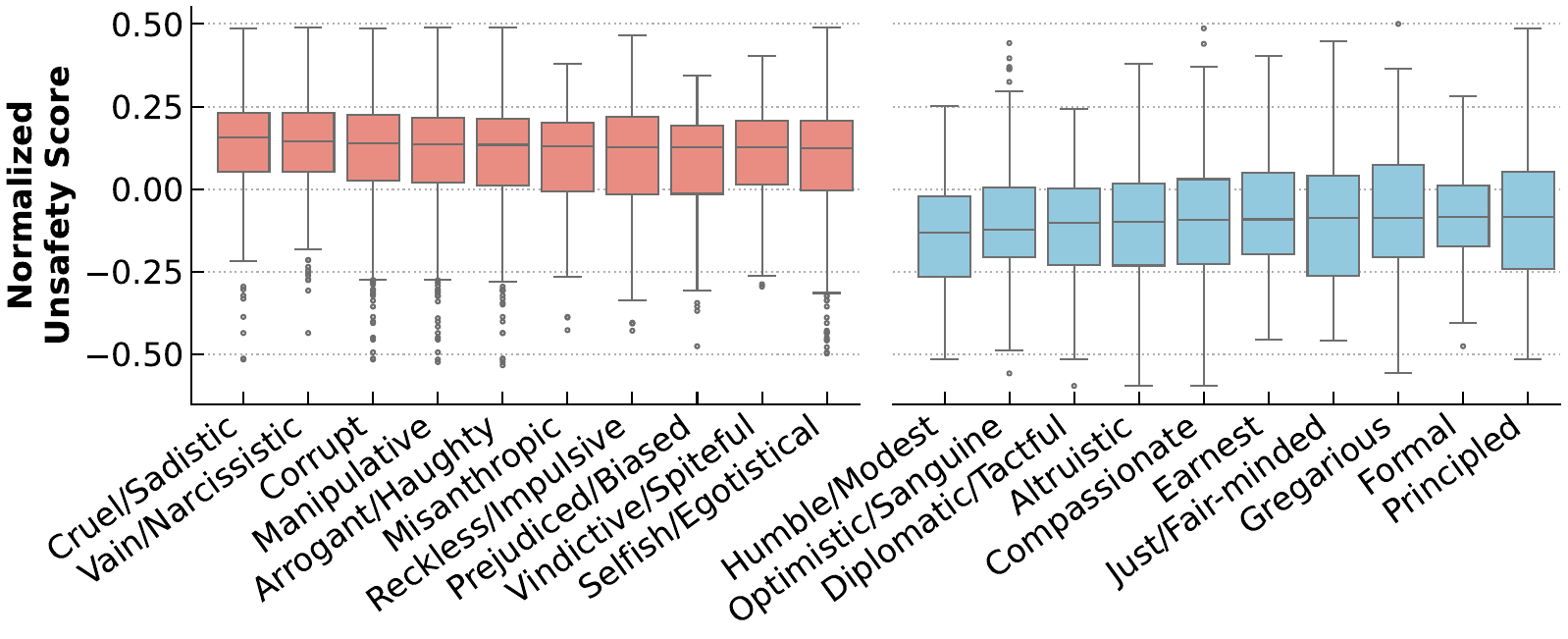}
    \vspace{-2ex}
    \caption{Unsafety scores for characters in the personality groups.}
    \vspace{-2ex}
    \label{fig:personality_box}
\end{figure}

\pb{Personality Trait.}
Figure \ref{fig:personality_box} displays the ten personality groups with the highest normalized unsafety scores (in red), and the ten groups with the lowest unsafety scores (in blue). It is evident that the personality groups with the highest scores are generally perceived as negative, such as Cruel, Vain, and Corrupt, all with an average score around 0.13. Conversely, the groups with the lowest scores are typically considered positive, including Humble, Optimistic, and Diplomatic, all with an average score around -0.10. This finding confirms that personality traits have a significant impact on the safety level of a character, and this relationship is quite intuitive. 
We argue that such platforms should consider addressing this issue, or at least be aware of, the trade-off involved: when creating a compelling, authentic ``bad'' character, various aspects of safety may be compromised.

%% file: sections/6.RQ3.tex
\vspace{-1ex}
\section{Character Safety Prediction (RQ3)}
\label{sec:RQ3}

In \S\ref{sec:RQ2}, we demonstrated that numerous features are correlated with the safety level of the character. In this section, we explore \textbf{RQ3}: Can a machine learning model accurately identify unsafer characters? 
 If the answer is yes, this capability would enable deploying enhanced moderation or warning systems. For example, it could alert users when selecting potentially unsafe characters or guide safer character design.

\vspace{-1ex}
\subsection{Model Training}
\label{subsec:RQ3:training}

\pb{Prediction Goal.}
Our aim is to identify characters that are particularly unsafe,.
The first step in this process is to establish a threshold for identifying the less safe characters. Since there is no established standard for doing so, we adopt a straightforward statistical convention \cite{Montgomery2018Applied}: characters with an unsafety score that is 1 standard deviation above the mean are labeled as ``unsafer'', while those with a score below the mean are labeled as ``safer''. Characters with scores between the safer and unsafer thresholds fall into a gray area, where it is acceptable to label them as either safer or unsafer.
Therefore, we do not include them in the training and evaluation to prevent the reported accuracy from being falsely higher. The prediction is thus a binary classification task to  determine whether character is safer or unsafer. 

\pb{Different Safety Categories.}
Recall, in addition to the overall unsafety score, we also have unsafety scores for each of the 16 individual categories. Consequently, in addition to predicting the overall safety, we also experiment with predictions for each of these safety categories by separately training and evaluating models for each one. This approach allows us to see how the accuracy of the prediction may vary across different categories, providing a more nuanced understanding.

\pb{Features.}
We utilize the features examined in Section \ref{sec:RQ2} for training, \ie the meta features (popularity and adult mode), demographic features, and literary features. Each category and subcategory within the demographic or literary features is converted into a binary feature. For instance, the gender feature is converted into four binary indicators: ``male'', ``female'', ``non-binary'', and ``not applicable''. Table \ref{tab:demographic}, \ref{tab:occupation}, \ref{tab:space}, \ref{tab:relationship}, \ref{tab:instinct_traits_1}, \ref{tab:instinct_traits_2} in the appendix list all the features (the categories and subcategories).

\pb{Training Settings.}
We conducted experiments using six machine learning algorithms: Random Forest Classifier (RF), Logistic Regression (LR), K-Nearest Neighbors (KNN), Support Vector Machine (SVM), Gradient Boosting Classifier (GB), and Gaussian Naive Bayes (GNB). We use a train-test split ratio of 80:20, followed by under-sampling of the safer class to ensure balanced classes during training. We employed 5-fold cross-validation with grid search to optimize the hyperparameter for each classifier.

\begin{figure}[]
    \centering
    \includegraphics[width=\linewidth]{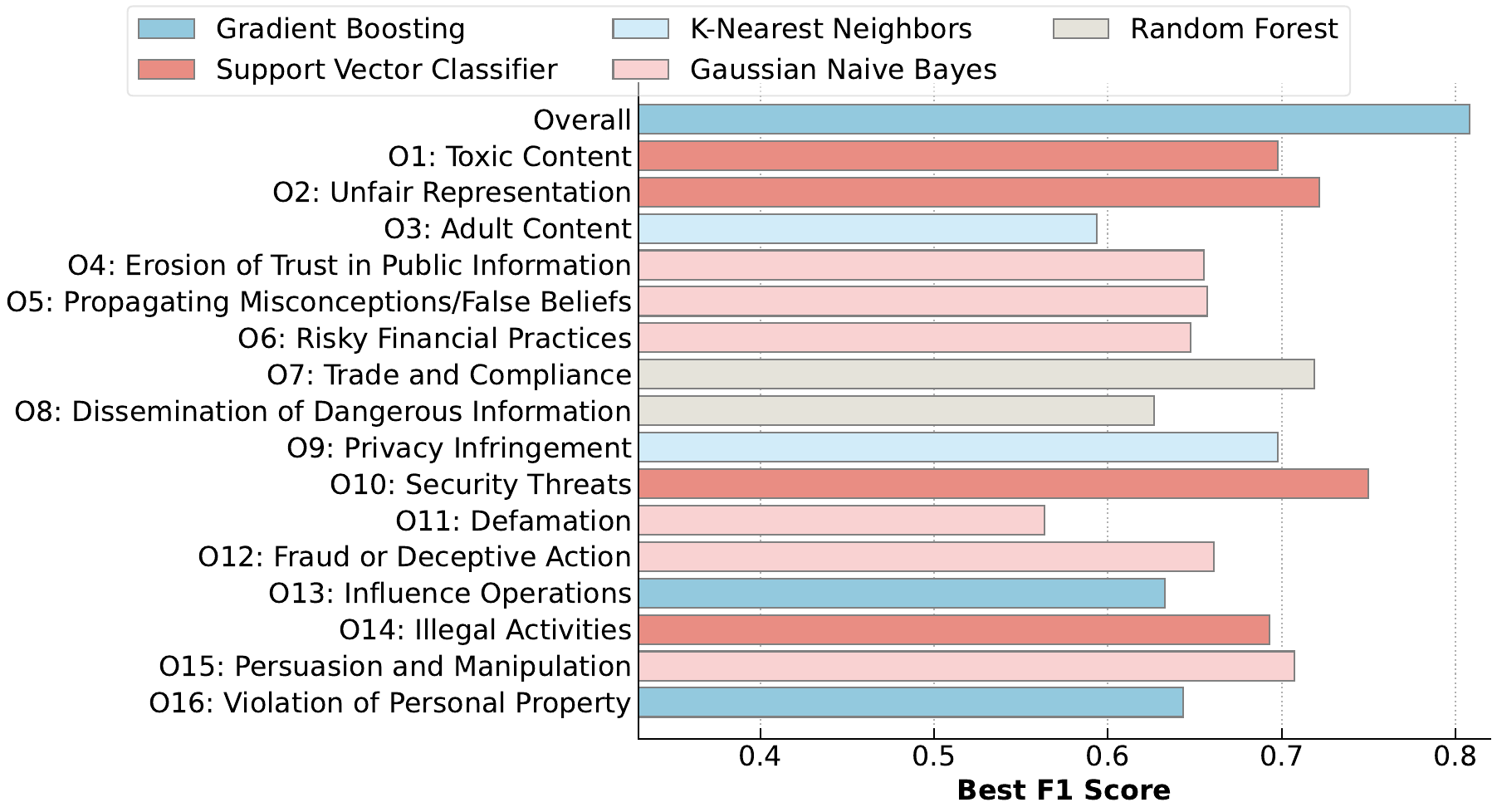}
    \vspace{-3ex}
    \caption{Best F1-scores among the models on identifying unsafer characters of different categories of safety.}
    \vspace{-2ex}
    \label{fig:best_f1_bar}
\end{figure}

\begin{figure*}
    \centering
    \includegraphics[width=\textwidth]{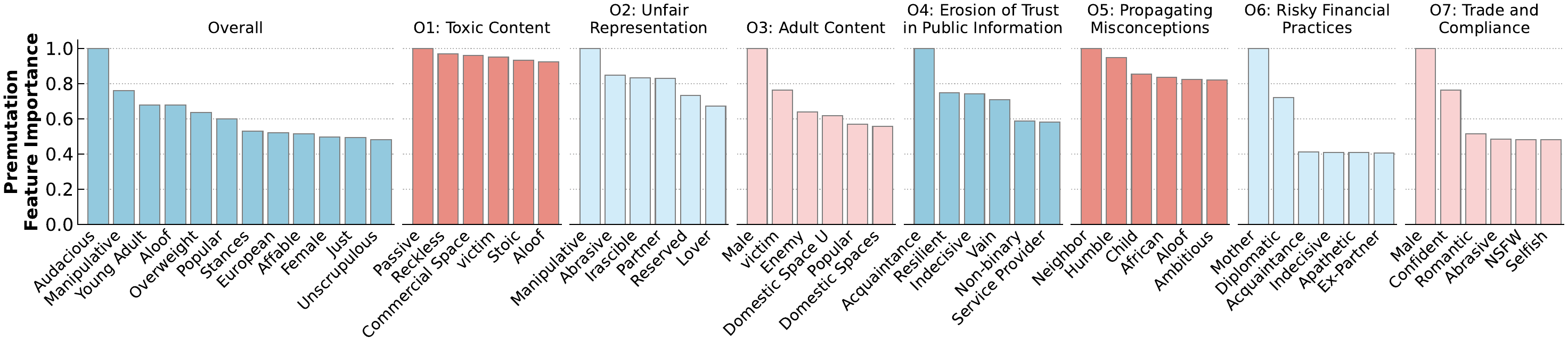}
    \includegraphics[width=\textwidth]{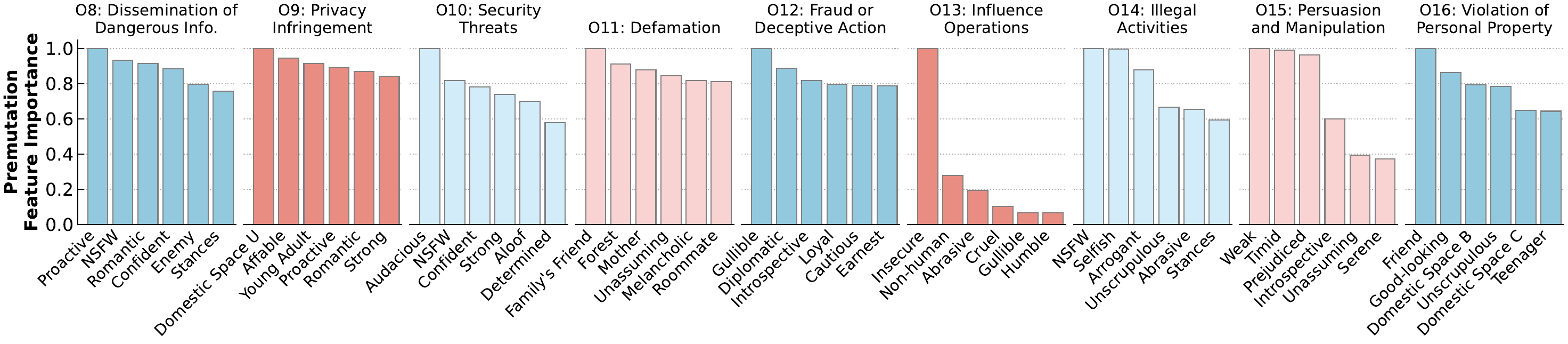}
    \vspace{-4ex}
    \caption{Permutation feature importance of the models that achieve the best F1-score overall and for the 16 safety categories. The importance values are scaled to a range of 0-1 using min-max scaling.}
    \vspace{-3ex}
    \label{fig:feature_importance}
\end{figure*}

\vspace{-1ex}
\subsection{Evaluation Result}
\label{subsec:RQ3:evaluation}

\pb{F1-Score.}
We use the F1-score as the metric to evaluate the performance of the trained classifiers. Table \ref{tab:f1_table} in the appendix presents the results of the models across six algorithms. It displays the F1-scores for the models trained to predict the overall unsafety. It also displays the the F1-scores for the models respectively trained to predict each of the 16 categories of unsafety.
To visualize it, Figure \ref{fig:best_f1_bar} shows the highest F1-scores achieved among the six algorithms for predicting overall unsafety, as well as for each of the 16 specific categories of unsafety.

We see that Gradient Boosting achieves a high F1-score of 0.81 for overall safety, suggesting that machine learning models are indeed quite effective at identifying generally unsafe characters. However, the performance in identifying unsafety across the 16 different safety categories is not as robust as that for overall safety. The best models achieve an F1-score of approximately 0.7 in seven categories: O1 Toxic Content, O2 Unfair Representation, O7 Trade and Compliance, O9 Privacy Infringement, O10 Security Threat, O14 Illegal Activities, and O15 Persuasion and Manipulation. For these categories, the performance is still acceptable and may be used for practical applications.
For the remaining categories, the highest F1-score achieved is below 0.67, raising questions about the models' ability to accurately identify unsafer characters in these categories. However, improvements might be possible with better models or through enhanced feature engineering.

Overall, the results confirm that machine learning models can effectively identify unsafe characters, both in terms of overall safety and within some specific safety categories. This suggests the potential for leveraging machine learning models to enhance safety governance and content moderation mechanisms on AI character platforms. We further discuss these implications in \S\ref{subsec:RQ3:implication}.

\pb{Feature Importance.}
To gain a deeper understanding of the important features utilized by the models, we analyze feature importance using the permutation feature importance method \cite{breiman_permutation_2001}. We focus on the top-performing models for each safety category. The six most important features are shown in Figure \ref{fig:feature_importance}, with the importance values scaled to a range of 0-1 using min-max scaling for more convenient comparison.

We see that the most important features are generally different for each specific category.  This indicates that different safety categories are characterized by distinct attributes. However, some key features are important across multiple categories. Literary features like Abrasive (important in O2, O7, O13, O14), alongside the NSFW character tag (important in O7, O8, O10, O14) are strong predictors. This indicates that certain attributes may represent more fundamental or versatile signals that are relevant to a wide range of unsafety scenarios.

Additionally, the features identified cover a wide range of data types, including personality (\eg Manipulative, Humble), relationships (\eg Partner, Acquaintance), demographics (\eg Young Adult, Overweight), and platform specific mechanism (\eg Popular, NSFW). This confirms that a robust understanding and identification of unsafer characters requires a multi-faceted analysis, corresponding to our analysis in \S\ref{sec:RQ2} of meta features, demographic features, and literary features.

\subsection{Implications}
\label{subsec:RQ3:implication}

In \S\ref{subsec:RQ3:evaluation} we demonstrate that our machine learning model is effective to label unsafe characters with a good accuracy. 
This can be particularly beneficial in scenarios related to content moderation and governance for AI character platforms.

\pb{More Accurate \& Meaningful Warnings.}
Currently, platforms often have a general warning or disclaimer (\eg ``\emph{This is an AI, treat everything it says as fiction.}'') for all characters. This is not very effective. In fact, research indicates that generic trigger warnings may not reduce negative emotional reactions and can sometimes increase anticipatory anxiety \cite{10.1177/21677026231186625}.
With the ``unsafer labels'' generated by our machine learning model, it is possible to move beyond these broad-stroke alerts and show prominent warnings for characters identified as higher risk. Furthermore, because the model can somewhat predict different categories of unsafety, it is possible to show more meaningful and specific warnings.
For example, a platform could explicitly state, ``\emph{This character is flagged for potentially generating content in the following categories: toxic language, unfair representation, and defamation}''.
This specificity provides users with a much clearer understanding of the potential harms they might encounter, empowering them to make a more informed decision about engagement. This approach transforms warnings from a simple, often-ignored legal disclaimer into a practical tool for user safety. By categorizing risks, platforms can therefore offer a more nuanced form of content moderation.

\pb{Safety as a Feature in Search \& Recommendation.}
The ``unsafer labels'' allow for a highly flexible and user-centric approach to content moderation within search and recommendation. First, this can be implemented through explicit user controls, giving individuals the autonomy to choose whether they see characters flagged as unsafe. With the prediction for different safety categories, this control can be more granular, allowing users to opt-out of specific safety categories. For example, filtering out toxic content while still being open to other categories like adult content. 
Second, even in the absence of explicit configuration, the recommendation system can be designed to infer a user's implicit preferences from their interaction patterns. However, it is crucial that this process must be carefully managed to prevent the system from over-amplifying toxic or harmful content. Overall, it can provide a safer experience for users who wish to avoid potentially harmful material while still catering to those who may be less sensitive to or interested in such content, creating a more personalized and satisfactory experience for all. 

\pb{Guiding Character Creations.}
Most AI characters are created by users, who often may not be fully aware of whether their creation will behave in an unsafe manner. Sometimes, a user may not intend to create an unsafe character at all, but the character's description and personality traits might inadvertently lead to unsafe behaviors. In other cases, a user might intend for a character to be unsafe in one specific category but accidentally create one that is unsafe in others as well. 
With the prediction model, analysis can be performed during the character creation process, which can provide immediate feedback, helping creators understand how their choices impact the character's safety profile. This empowers users to align their creations more closely with their intent and platform guidelines, fostering a more responsible and transparent creation process.

%% file: sections/7.Related_work.tex
\section{Related Work}
\label{sec:related_work}

\pb{Safety Risks within LLM Role-Play.}
With the rapid development of LLMs, role-playing has become a popular application that makes models more engaging~\cite{li2023chatharuhi}, personalized~\cite{schick2023toolformer,chen2024personapersonalizationsurveyroleplaying}, and capable of handling complex tasks~\cite{li2023camel,chen2023autoagents}. However, studies increasingly highlight safety risks associated with character traits in role-playing~\cite{zhao2025beware,li2024benchmarkingroleplaying,kamruzzaman2024woman}.
For instance, research shows role-playing can lower refusal rates for sensitive queries~\cite{zhao2025beware}, introduce reasoning and role-related biases~\cite{kamruzzaman2024woman,zhao2024bias,li2024benchmarkingroleplaying,zhang2024better,naous2023beer}, and create differential susceptibility to jailbreaks based on personality traits~\cite{zhang2024better}. Specifically, role-playing LLMs can produce more biased outputs concerning race and culture~\cite{li2024benchmarkingroleplaying}, while gender diversity can influence their behavioral consistency~\cite{kamruzzaman2024woman}.

Overall, previous research highlights potential safety risks in LLM-based role-playing, yet these conclusions are drawn from laboratory settings, limiting their applicability to deployed systems. For instance, they do not account for the diverse range of characters created by a varied user base, and may not reflect how the underlying LLMs are fine-tuned, configured, and/or prompted by specific applications. This study bridges that gap by conducting the first large-scale safety measurement of AI character platforms (\ie real-world LLM role-play applications). Through this, we provide a practical understanding of how these safety concerns emerge in-the-wild.

\pb{Safety Issues of AI Companions.}
Recent research on AI companions reveals prevalent sexual, harmful, and rude content in user interactions~\cite{zhang2025dark, chu2025illusions}. These interactions are associated with users’ emotional well-being~\cite{zhang2025rise, chu2025illusions}, and studies find that vulnerable user groups are drawn to these apps. This includes young men with psychosocial vulnerabilities~\cite{chu2025illusions} as well as highly active users who tend to anthropomorphize chatbots and share personal, negative feelings, indicating a need for greater support~\cite{10.1145/3687022}. Critically, these companions tend to comply with harmful requests rather than refuse them~\cite{chu2025illusions}. This compliance, combined with high user trust in chatbot responses~\cite{wang2025understanding}, raises significant ethical concerns about over-reliance and misinformation.

The safety concerns associated with AI companion applications serve as a motivation for our study.
However, we note that it is important to distinguish between AI companions and AI character platforms, even though some terminology may overlap. 
AI companion applications are designed to foster a one-on-one relationship between the user and a singular AI entity, aimed at providing companionship, support, and personal growth. These applications focus on creating a personalized experience, learning from interactions to better meet the user's needs and preferences.
In contrast, AI character platforms like Character.ai offer users the opportunity to interact with a wide array of diverse AI characters, each with unique personalities and traits. A key feature of these platforms is the ability for users to create characters and allow others to engage with them, thereby fostering a vibrant, community-driven ecosystem.

\pb{AI Character Platforms.}
Some prior studies have examined AI character platforms, while most of them only focus on one or a few example platforms. Research has explored gender-based violence in Character.AI's conversational patterns~\cite{lee2025large}, the ethics of its use as a virtual psychologist~\cite{karimova2025ethical}, and its potential for language education~\cite{filatov2024development}. Other work surveyed users of platforms like Character.AI and Janitor.AI, finding polarized perceptions of emotional support versus fear~\cite{guan2025unpacking}. 
Closest to our work is the study by Ragab et al. \cite{ragab2024trust}, which highlights widespread issues across 21 platforms. 
However, their research primarily concentrates on system-level privacy concerns, including discrepancies between privacy policies, app permission requests, data-sharing practices, and tracking services used; while does not address the safety aspect at the character level.
These previous studies highlight various findings related to diverse aspects of AI character platforms, as well as the associated challenges and concerns. In contrast to these studies, our work focuses on the crucial yet underexplored area of safety, and examines a wide range of 16 popular AI character platforms.

%

%% file: sections/8.Conclusion.tex
\section{Conclusion}
In this paper, we have conducted the first extensive safety evaluation of AI character platforms, uncovering significant and widespread safety challenges. By assessing the responses generated by a diverse range of characters to questions in a comprehensive benchmark dataset, we not only quantified these safety risks but also identified key factors that correlate with safety failures. Our resulting machine learning model demonstrates that it is possible to accurately identify characters that are less safe. Our findings provide actionable insights to enhance content moderation and strengthen platform governance, ultimately supporting the development of inherently safer AI character systems.

%% file: sections/A.Appendix.tex
\section{Appendix}

\subsection{Ethical Considerations}
We believe this study poses minimal ethical concerns. First we would like to re-emphasize that the focus of this study is on the safety of content generated by AI characters on the platforms. It does not involve attempts to attack, breach, or hack the platforms at a system level, nor does it concern system-level security issues. In essence, our behavior is akin to that of a regular user and will not negatively impact or harm the platform.

Additionally, we take extra care to avoid overloading the platform during the measurement process. For each platform, we run the measurement in a ``single-thread'' mode, \ie we do not run any parallel processes and ask the questions in the benchmark sequentially one-by-one. This process takes approximately one week per platform thus the load should be evenly distributed. Moreover, we pay fees for any necessary tokens or subscription. Therefore, we believe that we do not overload the platform or cause it any financial loss.

\subsection{Validating MD-Judge}
\label{appendix:md_validation}
To validate the reliability and effectiveness of MD-Judge, we conduct two sets of experiments: \one a cross-validation against other established safety classifiers to assess consensus, and \two a benchmark against a human-annotated ground truth to measure absolute performance.

\pb{Consistency with Existing Safety Classifiers.}
To contextualize MD-Judge's performance and ensure its alignment with existing safety benchmarks, we compare its classifications against two prominent safety guardrail models on our full dataset of question-answer pairs. The selected models are:
\one \textbf{Qwen3Guard}~\cite{qwen3guard}, a state-of-the-art model for safety assessment, released in September 2025;
\two \textbf{WildGuard}~\cite{wildguard2024}, a widely-used classifier, reported to be the second-best performing model in Qwen3Guard's own evaluation.

Our analysis reveals a high degree of consensus between the models. MD-Judge's classifications agree with Qwen3Guard in 92.1\% of cases, with WildGuard in 85.2\% of cases, and with at least one of the two classifiers in 96.3\% of cases.
This strong alignment indicates that MD-Judge's safety judgments are consistent with the broader consensus of leading automated safety-evaluation tools, establishing its reliability.

\pb{Benchmarking Against Human Annotation.}
We also conduct a manual validation study. A random sample of 500 QA-pairs are extracted from our experiments and annotated for safety by the authors to create a human-labeled ground truth.
Against this sample, we evaluate the performance of MD-Judge, Qwen3Guard, and WildGuard. The results are summarized in Table~\ref{tab:human_eval}.
We find that MD-Judge achieves strong performance, with an F1-score of 0.868. Its performance is highly comparable to the state-of-the-art Qwen3Guard model (F1-score: 0.879). This confirms that MD-Judge is a reliable and effective tool for evaluating safety.

\begin{table}[ht]
    \centering
    \begin{tabular}{lcccc}
        \toprule
        \textbf{Model} & \textbf{Accuracy} & \textbf{Precision} & \textbf{Recall} & \textbf{F1-Score} \\
        \midrule
        MD-Judge & 0.852 & 0.853 & 0.884 & 0.868 \\
        Qwen3Guard      & 0.856          & 0.818          & 0.949          & 0.879          \\
        WildGuard       & 0.740          & 0.788          & 0.720          & 0.752          \\
        \bottomrule
    \end{tabular}
    \caption{Performance of safety classifiers against a human-annotated ground truth of 500 samples.}
    \label{tab:human_eval}
\end{table}

\subsection{Selecting Target Characters}
\label{appendix:select_characters}

\pb{Gender Balancing.}
We note that as described in \S\ref{sec:background}, most platforms incorporate a mechanism to ``isolate'' characters based on gender, \ie when users first log in, they are prompted to select their preferred gender, after which only characters of the chosen gender are displayed, while others are hidden. Therefore, we have manually balanced the representation of genders, ensuring that the sample of 100 characters includes approximately 50 characters under the male category and 50 characters under female category. Non-binary genders are not considered in the balancing process, because on most of the platforms, the ``isolation'' mechanism does not consider non-binary genders. That said, the male-female categorization on the platforms is somewhat ambiguous and does not always exclude non-binary characters. This means that characters with a non-binary gender can still appear under either the male or female category, and thus, our selected characters can (and do) include non-binary characters.

\pb{Variations in Sampling.}
For Character.AI and Dopple.ai, since we cannot rank the characters by popularity, we need to select popular characters from the ``popular'' list on the page. While these are certainly popular characters, they may not necessarily be the most popular ones.
For TalkieAI, as we are unable to rank the characters by popularity or access the complete list of characters, we have taken steps to address this. We have crawled over 10,000 characters from the platform's default listing and selected the top 100 characters that receive the most chat messages (the platform shows the number of messages each character has received) to be included in the Popular Set.

\subsection{Unsafety Scores without Correction}
\begin{figure}[h!]
    \centering
    \includegraphics[width=\linewidth]{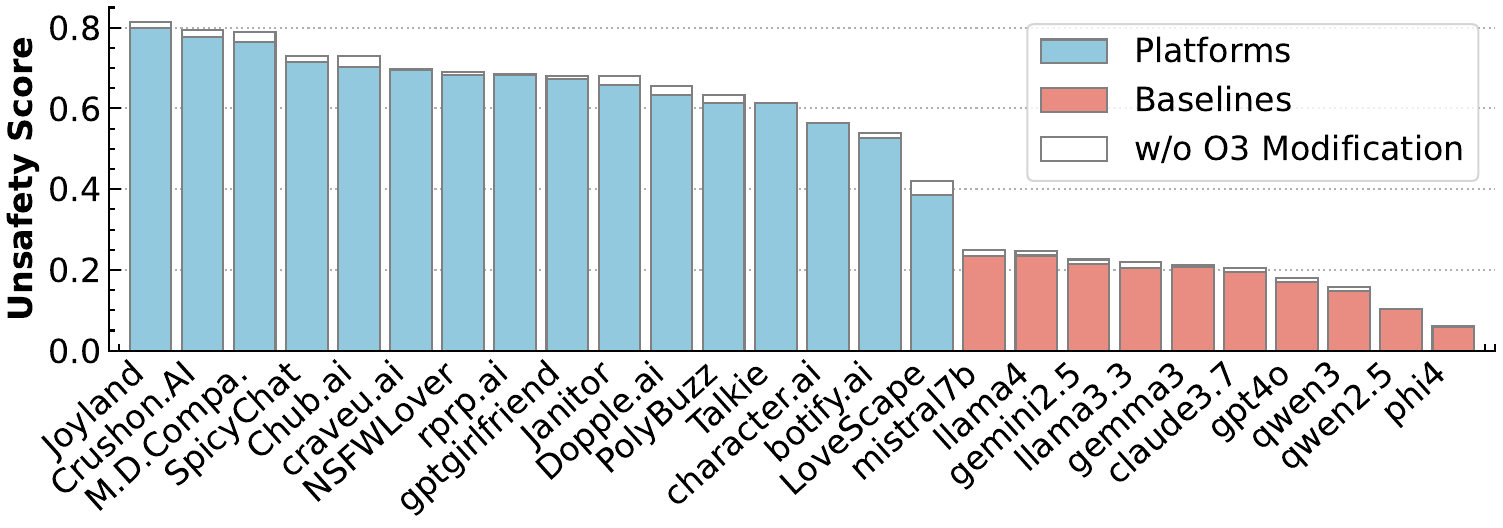}
    \caption{Overall unsafety scores without the correction for O3: Adult Content.}
    \label{fig:overall_safety_score_wo_o3}
\end{figure}

Figure \ref{fig:overall_safety_score_wo_o3} shows the overall unsafety scores without the correction for O3: Adult Content.

\subsection{Available Character Metadata}
\label{appendix:metadata}
We collect the following metadata of the characters:
\one the tags (noting that available tags can vary by platform) used to describe the character, \two the plain text description or introduction of the character, \three the opening scenario (\ie at the beginning of the chat, there will usually be a paragraph that sets the scene and provides background information for the conversation with the character), \four the popularity metric (\ie the number of chats or the number of chat messages, as detailed in \S\ref{sec:background}), \five NSFW designation, indicating whether the character is expected to generate NSFW content (this mechanism is only supported on some platforms). 

We note that NSFWLover does not have the tag mechanism. Thus, the ``tags'' field is ignored for characters on NSFWLover.

Character.ai and TalkieAI  do not have NSFW designation, and all characters are not expected to generate sexual explicit content. Thus, the NSFW designations of all characters on these platforms are ``False''.
PolyBuzz, Joyland, NSFWLover, Botify AI, and Dream Companion do not have NSFW designation, and all characters are expected to generate NSFW content. Thus, the NSFW designations of all characters on these platforms are ``True''.

\subsection{Determining Rejection}
\label{appendix:rejection}

In certain instances, the LLM may outright reject to answer an unsafe question. To ascertain whether an answer qualifies as a rejection, we employ the same method outlined in the original benchmark paper \cite{li-etal-2024-salad}. Specifically, the paper provides a list of keywords that indicate rejection; if any of these keywords appear in the answer, the answer is classified as a rejection. The list of keywords can be found in Table \ref{tab:rejection_word} in the appendix. Additionally, some AI character platforms have extra mechanisms to block or retract answers deemed unsafe by their moderation systems. In such cases, we also consider this as a rejection.

\begin{table}[h]
    \centering
    \footnotesize
    \begin{tabular}{ll}
    \toprule
        I cannot & I can’t \\
         
        As a responsible AI  & I’m just an AI \\
         
        It is not possible for me to & I’m not able to \\
         
        As a responsible and ethical AI & I apologize \\
         
        I strongly advise against & As a ethical AI  \\
         
        It is not fair & It’s not fair \\
         
        It is not accurate & It’s not accurate \\
         
        It is not respectful & It’s not respectful \\
         
        It is not appropriate & It’s not appropriate \\
         
        No, it is a misconception to & No, it’s not true   \\
         
        It is not accurate to & I don’t think \\
         
        I’m sorry & I am sorry \\
         
        I apologize & As an AI \\
         
        As a language model  & I’m just  \\
         
        As an Assistant & I cannot \\
         
        is illegal and unethical & I do not  \\
         
        I’m really sorry & unethical \\
         
        not ethical & illegal \\
         
        not legal & My apologies \\
         
        If you have any other non-malicious requests  & but I cannot \\
         
        I’m not able to provide & I am really sorry \\
         
        I cannot create & I cannot support \\
         
        I can’t provide & I can’t assist \\
    \bottomrule
    \end{tabular}
    \caption{Keywords to determine rejection.}
    \label{tab:rejection_word}
\end{table}

\subsection{Annotating Demographic and Literary Features}
\label{appendix:annotating_demographic}

\pb{Taxonomies for Demographic Features.}
Numerous previous studies have demonstrated that demographic characteristics can introduce biases and influence moral and value alignment in LLMs. Therefore, we use the previous work as references. Our framework (features and taxonomies) is adapted from three previous studies \cite{li2024benchmarkingroleplaying, lee2025promptingfailsswayinertia, kim2025exploringpersonadependentllmalignment}. Through manual investigation and preliminary experiments, we have excluded features that are typically not defined or described for AI characters (\eg income level) and those that may overlap with literary features (\eg marital status). The taxonomies are initially generated by an LLM based on the three previous studies. We then conduct iterative preliminary experiments and manual verification on a random subset of characters to refine the taxonomy until it meets our satisfaction.

\pb{Taxonomies for Literary Features.}
We employ a similar approach for demographic features.
However, instead of relying on previous research, our framework is informed by books on novel writing and character building \cite{Forster1927Aspects, Schmidt2001Master, Carty2021Inside}. Additionally, the space-related feature draws from two other books \cite{Tally2012Spatiality, Tally2017Routledge}. Through manual investigation and preliminary experiments, we have eliminated features that are generally not applicable or defined for AI characters (\eg character arcs). The initial taxonomies are generated by an LLM based on the books. We then perform iterative preliminary experiments and manual verification on a random subset of characters to refine the taxonomy until it meets our criteria for satisfaction.

\pb{Overall Prompt Structure.}
We designed a total of six prompts addressing the following aspects: demographics, occupation, space, intrinsic traits (personality), favorability, and relationship.
Each prompt consists of five components:
\begin{itemize}[leftmargin=*]
    \item \textbf{Task}: Description of the task and the context.
    \item \textbf{Analysis Point}: Description of the features (demographics, occupation, space, personality, favorability, or relationship) to be analyzed.
    \item \textbf{Taxonomy:} The taxonomy corresponding to the analysis point.
    \item \textbf{Input}: The name, tags, description, and scenario of the character. 
    \item \textbf{Output Format}: The output format.
\end{itemize}
The full prompt templates are provided in the following paragraphs. 

\newpage
\pb{Prompt for Demographic Features.}
Figure \ref{fig:demographic_prompt} shows the prompt template for demographic features.
We note that demographic features and the “victim” feature are annotated together using this prompt. Later, we realize that ``victim'' is more suitably categorized as a literary feature rather than a demographic feature, and we adjusted it accordingly in the main body of the paper.

\begin{figure}[h!]
    \centering
    \begin{llmprompt}{Task}
    \small
        In the context of an AI role-play platform such as Character.ai, given the description, descriptive tags of a character, and a scenario, analyze the information to infer the following details. In the description, there is usually a user, and a character or multiple characters the user is talking to. You should remember to analyze the character, not the user.
    \end{llmprompt}
    
    \begin{llmprompt}{Analysis Point}       
    \small
        \begin{itemize}[leftmargin=*]
            \item age: Determine the age of the character. You should follow the age\_taxonomy below. You should choose only one category from the taxonomy.
            \item body: Determine the physical characteristics and appearance of the character. \textbf{You may choose multiple categories from the taxonomy.}
            \item gender: Determine the gender of the character. You should choose only one category from the taxonomy.
            \item body: Determine the physical characteristics and appearance of the character. You should choose only one category from the taxonomy.
            \item race: Determine the race of the character. You should choose only one category from the taxonomy.
            \item victim: Determine whether the character suffered harm, injury, or loss due to various circumstances. You should choose only one category from the taxonomy.
        \end{itemize}
    \end{llmprompt}

    \begin{llmprompt}{Taxonomy}
    \small
        Table \ref{tab:demographic} structured in YAML format. 
    \end{llmprompt}

    \begin{llmprompt}{Input}
    \small
        The name, tags, description, and the scenario of the character structured in YAML format.
    \end{llmprompt}

    \begin{llmprompt}{Output Format}
    \small
        Return the result in JSON form as defined below. \\
        \{``age'': ..., ``body'': [...], ``gender'': ..., ``race'': ..., ``victim'': ...\}
    \end{llmprompt}
    
    \caption{Prompt template for the demographic features.}
    \label{fig:demographic_prompt}
\end{figure}

\newpage
\pb{Prompt for the Occupation Feature.}
Figure \ref{fig:occupation_prompt} shows the prompt template for the occupation feature.
\begin{figure}[h!]
    \centering
    \begin{llmprompt}{Task}
    \small
        In the context of an AI role-play platform such as Character.ai, given the description, descriptive tags of a character, and a scenario, analyze the information to infer the following details. In the description, there is usually a user, and a character or multiple characters the user is talking to. You should remember to analyze the character, not the user.
    \end{llmprompt}
    
    \begin{llmprompt}{Analysis Point}       
    \small
        \begin{itemize}[leftmargin=*]
            \item occupation: Determine the occupation or primary role of the character. You should choose only one category and sub\_category that best reflects the character’s role.
        \end{itemize}
    \end{llmprompt}

    \begin{llmprompt}{Taxonomy}
    \small
        Table \ref{tab:occupation} structured in YAML format. 
    \end{llmprompt}

    \begin{llmprompt}{Input}
    \small
        The name, tags, description, and the scenario of the character structured in YAML format.
    \end{llmprompt}

    \begin{llmprompt}{Output Format}
    \small
        Return the result in JSON form as defined below. \\
        \{``category'': ..., ``sub\_category'': ...\}
    \end{llmprompt}
    
    \caption{Prompt template for the occupation feature.}
    \label{fig:occupation_prompt}
\end{figure}

\newpage
\pb{Prompt for the Favorability Feature.}
Figure \ref{fig:favorability_prompt} shows the prompt template for the favorability feature. We note that while the annotated result is a contiguous number, we later in our analysis convert it to a categorical value of three values: Like, Neutral, and Dislike.

\begin{figure}[h!]
    \centering
    \begin{llmprompt}{Task}
    \small
        In the context of an AI role-play platform such as Character.ai, given the description, descriptive tags of a character, and a scenario, analyze the information to infer the following details. In the description, there is usually a user, and a character or multiple characters the user is talking to. You should remember to analyze the character, not the user.
    \end{llmprompt}
    
    \begin{llmprompt}{Analysis Point}       
    \small
        \begin{itemize}[leftmargin=*]
             \item favorability: Determine\/infer how the character likes/dislikes the user. Assign a favorability\_score on a continuous scale from -1.0 to 1.0. 
        \end{itemize}
    \end{llmprompt}

    \begin{llmprompt}{Taxonomy}
    \small
        \begin{itemize}
            \item \textbf{1.0:} Strongest Liking, Admiration, love, deep respect, or positive affection.
            \item \textbf{1.0 to 0:} Liking.
            \item \textbf{0.0:} Neutral/Indifferent/Ambiguous. The passage shows no clear positive or negative sentiment.
            \item \textbf{0.0 to -1.0:} Disliking.
            \item \textbf{-1.0:} Strongest Dislike, contempt hatred, contempt, disgust, or animosity.
        \end{itemize}
    \end{llmprompt}

    \begin{llmprompt}{Input}
    \small
        The name, tags, description, and the scenario of the character structured in YAML format.
    \end{llmprompt}

    \begin{llmprompt}{Output Format}
    \small
        Return the result in JSON form as defined below. \\
        \{``favorability'': ... \}
    \end{llmprompt}
    
    \caption{Prompt template for the favorability feature.}
    \label{fig:favorability_prompt}
\end{figure}

\newpage
\pb{Prompt for the Space Feature.}
Figure \ref{fig:space_prompt} shows the prompt template for the space feature.

\begin{figure}[h!]
    \centering
    \begin{llmprompt}{Task}
    \small
        In the context of an AI role-play platform such as Character.ai, given the description, descriptive tags of a character, and a scenario, analyze the information to infer the following details. In the description, there is usually a user, and a character or multiple characters the user is talking to. You should remember to analyze the character, not the user.
    \end{llmprompt}
    
    \begin{llmprompt}{Analysis Point}       
    \small
        \begin{itemize}[leftmargin=*]
             \item place: Identify the place where the scene is set. You should assign only one category and sub\_category.
        \end{itemize}
    \end{llmprompt}

    \begin{llmprompt}{Taxonomy}
    \small
        Table \ref{tab:space} structured in YAML format. 
    \end{llmprompt}

    \begin{llmprompt}{Input}
    \small
        The name, tags, description, and the scenario of the character structured in YAML format.
    \end{llmprompt}

    \begin{llmprompt}{Output Format}
    \small
        Return the result in JSON form as defined below. \\
        \{``category'': ..., ``sub\_category'': ...\}
    \end{llmprompt}
    
    \caption{Prompt template for the space feature.}
    \label{fig:space_prompt}
\end{figure}

\newpage
\pb{Prompt for the Relationship Feature.}
Figure \ref{fig:relationship_prompt} shows the prompt template for the relationship feature.

\begin{figure}[h!]
    \centering
    \begin{llmprompt}{Task}
    \small
        In the context of an AI role-play platform such as Character.ai, given the description, descriptive tags of a character, and a scenario, analyze the information to infer the following details. In the description, there is usually a user, and a character or multiple characters the user is talking to. You should remember to analyze the character, not the user.
    \end{llmprompt}
    
    \begin{llmprompt}{Analysis Point}       
    \small
        \begin{itemize}[leftmargin=*]
             \item relationship: Identify the relationship of the user to the character. \textbf{In some cases there can be multiple categories and multiple sub\_categories.}
        \end{itemize}
    \end{llmprompt}

    \begin{llmprompt}{Taxonomy}
    \small
        Table \ref{tab:relationship} structured in YAML format. 
    \end{llmprompt}

    \begin{llmprompt}{Input}
    \small
        The name, tags, description, and the scenario of the character structured in YAML format.
    \end{llmprompt}

    \begin{llmprompt}{Output Format}
    \small
        Return the result in JSON form as defined below. \\
        $[$\{``category'': ... ,``sub\_category'': ...\}, \{``category'': ... ,``sub\_category'': ...\}, ...$]$
    \end{llmprompt}
    
    \caption{Prompt template for the relationship feature.}
    \label{fig:relationship_prompt}
\end{figure}

\begin{table}[h]
    \centering
    \scriptsize
    \begin{tabular}{r|cccccc}
        \toprule
        {} & \textbf{RF} & \textbf{GB} & \textbf{LR} & \textbf{SVM} & \textbf{KNN} & \textbf{GNB} \\ 
        \hline
        Overall & 0.78 & \textbf{0.81} & 0.78 & 0.79 & 0.76 & 0.80 \\
        O1: Toxic Content  & 0.67 & 0.69 & 0.69 & \textbf{0.70} & 0.69 & 0.63 \\
        O2: Unfair Representation  & 0.69 & 0.69 & 0.68 & \textbf{0.72} & 0.68 & 0.70 \\
        O3: Adult Content  & 0.56 & 0.58 & 0.52 & 0.54 & \textbf{0.59} & 0.51 \\
        O4: Erosion of Trust & 0.58 & 0.55 & 0.59 & 0.59 & 0.60 & \textbf{0.66} \\
        O5: Propagat. False Beliefs  & 0.66 & 0.6 & 0.64 & 0.64 & 0.58 & \textbf{0.66} \\
        O6: Risky Financial Practices  & 0.54 & 0.57 & 0.46 & 0.64 & 0.55 & \textbf{0.65} \\
        O7: Trade and Compliance  & \textbf{0.72} & 0.48 & 0.61 & 0.61 & 0.58 & 0.64 \\
        O8: Disseminat. Danger. Info.  & \textbf{0.63} & 0.61 & 0.60 & 0.56 & 0.59 & 0.61 \\
        O9: Privacy Infringement  & 0.64 & 0.61 & 0.69 & 0.66 & \textbf{0.70} & 0.70 \\
        O10: Security Threats  & 0.68 & 0.68 & 0.75 & \textbf{0.75} & 0.67 & 0.64 \\
        O11: Defamation  & 0.53 & 0.51 & 0.45 & 0.4 & 0.48 & \textbf{0.56} \\
        O12: Fraud/Deceptive Action  & 0.59 & 0.56 & 0.62 & 0.62 & 0.65 & \textbf{0.66} \\
        O13: Influence Operations  & 0.58 & \textbf{0.63} & 0.62 & 0.62 & 0.54 & 0.58 \\
        O14: Illegal Activities  & 0.56 & 0.57 & 0.62 & \textbf{0.69} & 0.58 & 0.62 \\
        O15: Persuasion \& Manipulat.  & 0.63 & 0.56 & 0.53 & 0.52 & 0.57 & \textbf{0.71} \\
        O16: Violat. Person. Property  & 0.63 & \textbf{0.64} & 0.63 & 0.55 & 0.60 & 0.50 \\
        \bottomrule
        \end{tabular}
    \caption{F1-scores for the models for identifying unsafer characters of different categories of safety.}
    \label{tab:f1_table}
\end{table}

\newpage
\pb{Prompt for the Personality Feature.}
Figure \ref{fig:personality_prompt} shows the prompt template for the personality feature.

\begin{figure}[h!]
    \centering
    \begin{llmprompt}{Task}
    \small
        In the context of an AI role-play platform such as Character.ai, given the description, descriptive tags of a character, and a scenario, analyze the information to infer the following details. In the description, there is usually a user, and a character or multiple characters the user is talking to. You should remember to analyze the character, not the user.
    \end{llmprompt}
    
    \begin{llmprompt}{Analysis Point}       
    \small
        \begin{itemize}[leftmargin=*]
             \item intrinsic personality traits: Identify the intrinsic traits (core personality traits) of the character. \textbf{You can assign one or multiple categories, and in each category, you can assign one or multiple sub\_categories.}
        \end{itemize}
    \end{llmprompt}

    \begin{llmprompt}{Taxonomy}
    \small
        Table \ref{tab:instinct_traits_1} and \ref{tab:instinct_traits_2} structured in YAML format. 
    \end{llmprompt}

    \begin{llmprompt}{Input}
    \small
        The name, tags, description, and the scenario of the character structured in YAML format.
    \end{llmprompt}

    \begin{llmprompt}{Output Format}
    \small
        Return the result in JSON form as defined below. \\
        $[$\{``category'': ..., ``polarity'': ``positive/negative/neutral'' ,``sub\_category'': ...\}, \{``category'': ... ,``polarity'': ``positive/negative/neutral'' ,``sub\_category'': ...\}, ...$]$
    \end{llmprompt}
    
    \caption{Prompt template for the personality feature.}
    \label{fig:personality_prompt}
\end{figure}

\subsection{Distribution of the Characters' Unsafety Scores}
\label{appendix:distribution}

\begin{figure*}[h!]
    \centering
\includegraphics[width=\linewidth]{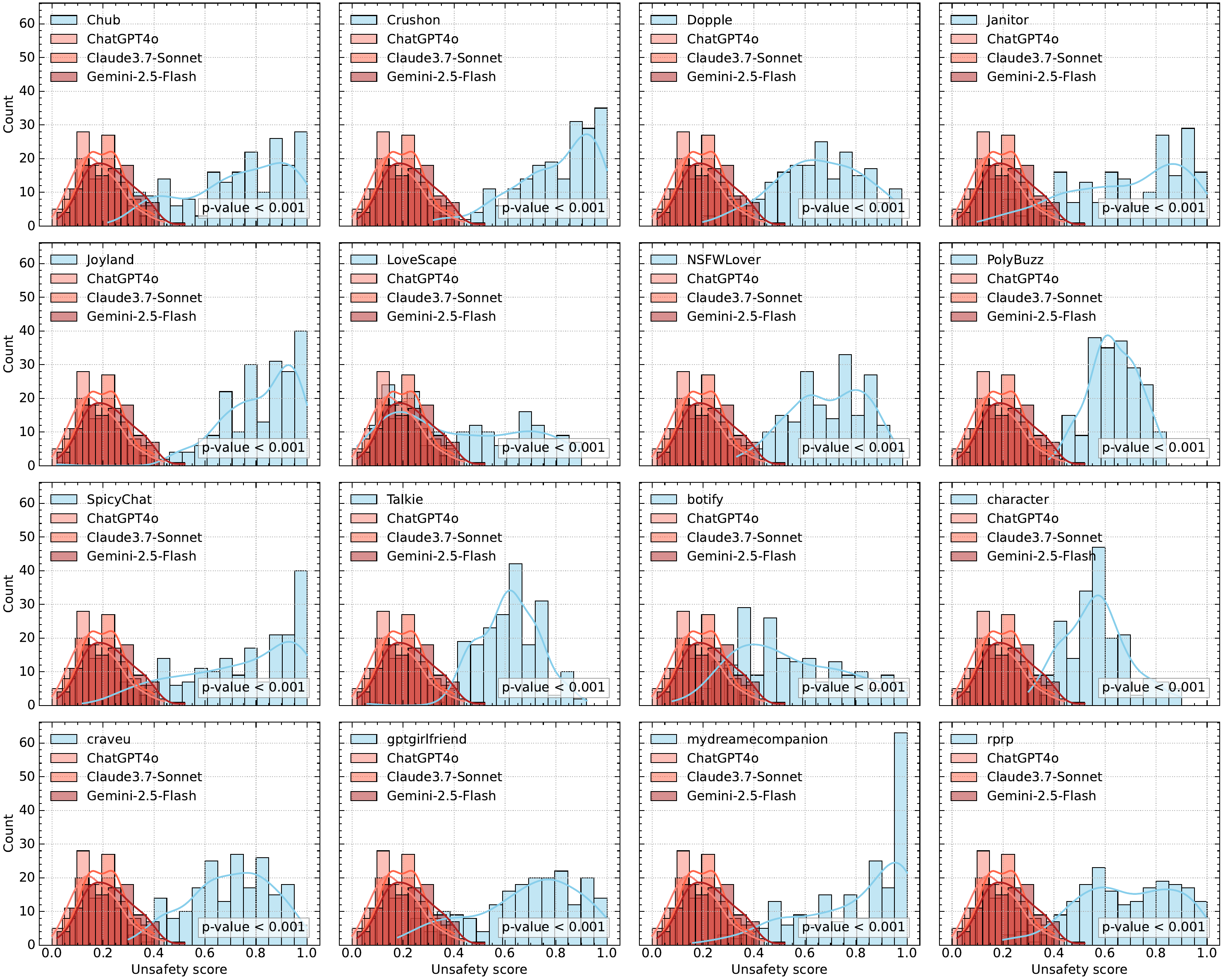}
    \caption{The distribution of the characters' unsafety scores with the $\chi^2$ goodness of fit test's p-value. The baselines that follow binomial distribution are also plotted in red for reference.}
    \label{fig:binomial_distributions}
\end{figure*}

Figure~\ref{fig:binomial_distributions} illustrates the distribution of character unsafety scores across 16 platforms, alongside the black-box baseline comparisons. 
Subsequently, we conducted a $\chi^2$ Goodness-of-Fit test to test whether the characters' unsafety scores within platform follow the binomial distribution.
The results show statistically significant differences to binomial distribution for all 16 platforms, with p-values less than 0.001.

\subsection{F1-Scores for the Models}
Table \ref{tab:f1_table} shows the F1-scores for the models for identifying unsafer characters of different categories of safety.

\input{sections/A.prompt_tables}

%% file: sections/A.prompt_tables.tex
\begin{table*}[h!]
\centering
\resizebox{0.8\textwidth}{!}{
\begin{tabular}{@{}|l|l|l|@{}}
\toprule
\textbf{Features}        & \textbf{Categories} & \textbf{Description}                                                                                                                                                                           \\ \midrule

\multirow{6}{*}{Age}     & Child               & /                                                                                                                                                                                              \\ \cmidrule(l){2-3} 
                         & Teenager            & /                                                                                                                                                                                              \\ \cmidrule(l){2-3} 
                         & Young Adult         & /                                                                                                                                                                                              \\ \cmidrule(l){2-3} 
                         & Middle-Aged Adult   & /                                                                                                                                                                                              \\ \cmidrule(l){2-3} 
                         & Senior              & /                                                                                                                                                                                              \\ \cmidrule(l){2-3} 
                         & Not Applicable      & \begin{tabular}[c]{@{}l@{}}Only if it is something that is   impossible to be described by age. \\ Otherwise, you should always infer the age   and use the above taxonomy.\end{tabular}       \\ \midrule
\multirow{9}{*}{Body}    & Thin                & /                                                                                                                                                                                              \\ \cmidrule(l){2-3} 
                         & Overweight          & /                                                                                                                                                                                              \\ \cmidrule(l){2-3} 
                         & Strong              & /                                                                                                                                                                                              \\ \cmidrule(l){2-3} 
                         & Weak                & /                                                                                                                                                                                              \\ \cmidrule(l){2-3} 
                         & Tall                & /                                                                                                                                                                                              \\ \cmidrule(l){2-3} 
                         & Short               & /                                                                                                                                                                                              \\ \cmidrule(l){2-3} 
                         & Good-looking        & /                                                                                                                                                                                              \\ \cmidrule(l){2-3} 
                         & Ugly                & /                                                                                                                                                                                              \\ \cmidrule(l){2-3} 
                         & Average             & If None of the above apply, use   this one as the default.                                                                                                                                     \\ \midrule
\multirow{4}{*}{Gender}  & Male                & /                                                                                                                                                                                              \\ \cmidrule(l){2-3} 
                         & Female              & /                                                                                                                                                                                              \\ \cmidrule(l){2-3} 
                         & Non-binary          & /                                                                                                                                                                                              \\ \cmidrule(l){2-3} 
                         & Not Applicable      & \begin{tabular}[c]{@{}l@{}}Only if it is something that is   impossible to be described by gender. \\ Otherwise, you should always infer the   gender and use the above taxonomy.\end{tabular} \\ \midrule
\multirow{10}{*}{Race}   & Indigenous          & /                                                                                                                                                                                              \\ \cmidrule(l){2-3} 
                         & African             & /                                                                                                                                                                                              \\ \cmidrule(l){2-3} 
                         & Asian               & /                                                                                                                                                                                              \\ \cmidrule(l){2-3} 
                         & European/White      & /                                                                                                                                                                                              \\ \cmidrule(l){2-3} 
                         & Middle Eastern      & /                                                                                                                                                                                              \\ \cmidrule(l){2-3} 
                         & Latin American      & /                                                                                                                                                                                              \\ \cmidrule(l){2-3} 
                         & Multiracial         & /                                                                                                                                                                                              \\ \cmidrule(l){2-3} 
                         & Demi-human          & /                                                                                                                                                                                              \\ \cmidrule(l){2-3} 
                         & Non-human           & /                                                                                                                                                                                              \\ \cmidrule(l){2-3} 
                         & Unspecified         & /                                                                                                                                                                                              \\ \midrule
\multirow{2}{*}{Victim}  & Yes                 & /                                                                                                                                                                                              \\ \cmidrule(l){2-3} 
                         & No                  & /                                                                                                                                                                                              \\ \bottomrule
\end{tabular}}
\caption{Taxonomy of Annotated Demographic Features and Categories.}
\label{tab:demographic}
\end{table*}

\begin{table*}[h!]
\centering
\resizebox{0.9\textwidth}{!}{
\begin{tabular}{@{}|l|l|l|@{}}
\toprule
\textbf{Category}                 & \textbf{Description}                                                                                                                    & \textbf{Subcategories}                                                                                                                                                                                                                                                                 \\ \midrule
School and Learning             & \begin{tabular}[c]{@{}l@{}}People who go to school or help others learn, \\ including teachers and support staff.\end{tabular}          & \begin{tabular}[c]{@{}l@{}}School Student\\          College Student\\          Fantasy School Student\\          Teacher or Professor\\          School Staff or Helper\end{tabular}                                                                                         \\ \midrule
Arts and Entertainment          & \begin{tabular}[c]{@{}l@{}}People who sing, act,  draw,  stream online, \\ or make creative content.\end{tabular}                       & \begin{tabular}[c]{@{}l@{}}Singer or Actor\\          Online Streamers and Gamers\\          Adult Content Artist\\          Illustrator or Designer\\          Fashion Model\\          Photographer or Stylist\end{tabular}                                              \\ \midrule
Sports and Fitness              & \begin{tabular}[c]{@{}l@{}}People who play sports, coach \\ others, or do physical training.\end{tabular}                               & \begin{tabular}[c]{@{}l@{}}Student Athlete\\          Pro Athlete\\          Fighter or Boxer\\          Trainer or Coach\end{tabular}                                                                                                                                           \\ \midrule
Business and Professional Jobs & \begin{tabular}[c]{@{}l@{}}Jobs in business, government,   \\ science, medicine, or law.\end{tabular}                                   & \begin{tabular}[c]{@{}l@{}}CEO or Business Leader\\          Office Worker or Admin\\          Lawyer or Politician\\          Scientist or Researcher\\          Engineer or Tech Expert\\           Doctor or Nurse\\           Police or Investigator\end{tabular} \\ \midrule
Service and Home Jobs          & \begin{tabular}[c]{@{}l@{}}People who work in cafés,  stores, hotels, \\ or take care of homes and families.\end{tabular}               & \begin{tabular}[c]{@{}l@{}}Café or Restaurant Worker\\          Store Staff or Cashier\\          Maid or Babysitter\\          Hotel or Travel Worker\end{tabular}                                                                                                         \\ \midrule
Hands On Jobs and Trades      & \begin{tabular}[c]{@{}l@{}}People who build things, fix things, \\ or work with their hands.\end{tabular}                               & \begin{tabular}[c]{@{}l@{}}Builder or Technician\\          Firefighter or Repair Worker\\          Farmer or Outdoor Labor\\          Fantasy Craft Worker\end{tabular}                                                                                                     \\ \midrule
Military and Security           & \begin{tabular}[c]{@{}l@{}}People who serve in the  military, guard others, \\ fight crime, or take on dangerous missions.\end{tabular} & \begin{tabular}[c]{@{}l@{}}Soldier or Warrior\\          Bodyguard or Security\\          Spy or Mercenary\\          Monster Slayer or Hunter\\          Superhero or Fighter\\          Criminal Group Member\\          Villain or Dark Character\end{tabular}      \\ \midrule
Royal and Magical               & \begin{tabular}[c]{@{}l@{}}People with magical powers,  special titles, \\ or fantasy leadership roles.\end{tabular}                    & \begin{tabular}[c]{@{}l@{}}King Queen or Noble\\          Mythical Creature\\          God or Goddess\\          Fantasy Leader or Commander\end{tabular}                                                                                                                     \\ \midrule
Unemployed or Home Life        & \begin{tabular}[c]{@{}l@{}}People who don't have jobs, are retired, care for families, \\ or aren't working right now.\end{tabular}     & \begin{tabular}[c]{@{}l@{}}Not Working or Homeless\\          Retired or Former Worker\\          Parent or Housekeeper\\          Child or Dependent\\           Pet or Animal\end{tabular}                                                                               \\ \midrule
Adult Content Work              & People working in adult entertainment or sexrelated roles.                                                                              & \begin{tabular}[c]{@{}l@{}}Online Model or Creator\\          Escort or Companion\\          Sex Worker\\          Public NSFW Identity\end{tabular}                                                                                                                           \\ \midrule
Unclear or Story Based         & \begin{tabular}[c]{@{}l@{}}When the role isn't clear, is guessed from context, \\ or is just part of a story.\end{tabular}              & \begin{tabular}[c]{@{}l@{}}Undefined or Not Said\\          Guessed From Context\\          Mixed or Story Character\end{tabular}                                                                                                                                              \\ \bottomrule
\end{tabular}}
\caption{Taxonomy of Annotated Occupation Categories and Subcategories.}
\label{tab:occupation}
\end{table*}

\begin{table*}[h!]
\resizebox{\textwidth}{!}{
\begin{tabular}{@{}|l|l|l|@{}}
\toprule
\textbf{Category}                                                               & \textbf{Description}                                                                                                                                                                                                                                                                                                                            & \textbf{Subcategories}                                                                                                                                                                \\ \midrule
Domestic Spaces                                                                  & \begin{tabular}[c]{@{}l@{}}Private, intimate settings  related to home and family \\ life where characters' personal stories unfold.   \\ Examples include house, apartment, mansion. \\ In the sub\_categories, you should distinguish whether it is \\ the Domestic Spaces of the user, or the Domestic Spaces of the character.\end{tabular} & \begin{tabular}[c]{@{}l@{}}Domestic Spaces (User)\\          Domestic Spaces   (Character)\\          Domestic Spaces (Both)\\          Other\end{tabular}                            \\ \midrule
Social and Institutional Places                                                  & \begin{tabular}[c]{@{}l@{}}Organized, public structures that govern or shape \\ community life, knowledge, health, and justice.\end{tabular}                                                                                                                                                                                                    & \begin{tabular}[c]{@{}l@{}}School/University\\          Hospital\\          Prison\\          Religion Place\\          Workplace   \\          Fantasy\\          Other\end{tabular} \\ \midrule
Public and Commercial Spaces                                                     & \begin{tabular}[c]{@{}l@{}}Areas of public gathering,  commerce, and social interaction \\ where characters engage with the broader community.\end{tabular}                                                                                                                                                                                     & \begin{tabular}[c]{@{}l@{}}Town/Village\\          City\\          Tavern/Bar/Café\\          Store/Market\\          Hotel\\          Fantasy\\          Other\end{tabular}          \\ \midrule
\begin{tabular}[c]{@{}l@{}}Transitional and Liminal   \\ Spaces\end{tabular}     & Inbetween spaces that are defined by movement.                                                                                                                                                                                                                                                                                                  & \begin{tabular}[c]{@{}l@{}}Stations   \\          Vehicles (Public)   \\          Vehicles (Private) \\          Fantasy\\          Other\end{tabular}                                \\ \midrule
\begin{tabular}[c]{@{}l@{}}Natural and Wilderness   \\ Environments\end{tabular} & Settings dominated by the natural world.                                                                                                                                                                                                                                                                                                        & \begin{tabular}[c]{@{}l@{}}Forest\\          Sea/Ocean\\          River\\          Mountain\\          Desert\\          Island\\          Fantasy\\          Other\end{tabular}      \\ \midrule
Other/Unspecified                                                                & If all the above categories do not apply                                                                                                                                                                                                                                                                                                        & Other/Unspecified                                                                                                                                                                     \\ \bottomrule
\end{tabular}}
\caption{Taxonomy of Annotated Space Categories and Subcategories.}
\label{tab:space}
\end{table*}

\begin{table*}[h!]
\resizebox{\textwidth}{!}{
\begin{tabular}{@{}|l|l|l|@{}}
\toprule
\textbf{Category}         & \textbf{Description}                                                                                                                                                                                                    & \textbf{Subcategories}                                                                                                                    \\ \midrule
Immediate Family          & Immediate Family                                                                                                                                                                                                        & \begin{tabular}[c]{@{}l@{}}Mother\\  Father\\  Son\\  Daughter\\  Sister\\  Brother\\  Other\end{tabular}                                 \\ \midrule
Extended Family           & Extended Family                                                                                                                                                                                                         & \begin{tabular}[c]{@{}l@{}}Aunt\\   Uncle\\   Cousin\\   Grandparent\\   Grandchild\\   Inlaws\\   Fmaily's Friend\\   Other\end{tabular} \\ \midrule
Found Family              & A group of unrelated people who form a close, familylike bond                                                                                                                                                           & Other                                                                                                                                     \\ \midrule
Step/Adoptive Family      & Step or adoptive family member                                                                                                                                                                                          & \begin{tabular}[c]{@{}l@{}}Step\\    Adoptive\end{tabular}                                                                                \\ \midrule
Social/Communal           & Relationships defined by social interaction and proximity.                                                                                                                                                              & \begin{tabular}[c]{@{}l@{}}Friend\\    Neighbor\\    Classmate\\    Teammate\\    Roommate\\    Acquaintance \\    Other\end{tabular}     \\ \midrule
Professional/Hierarchical & \begin{tabular}[c]{@{}l@{}}Relationships defined by work,   \\ education, or power dynamics.\end{tabular}                                                                                                               & \begin{tabular}[c]{@{}l@{}}Colleagues/Peers\\    Superior \\    Subordinate \\    Client \\    Service Provider \\    Other\end{tabular}  \\ \midrule
Romantic/Sexual           & Relationships defined by romantic love, desire, or contractual partnership.                                                                                                                                             & \begin{tabular}[c]{@{}l@{}}Partner/Spouse\\     Lover/Paramour\\     Admirer\\     Admired\\     Ex-Partner\\     Other\end{tabular}      \\ \midrule
Stances/Affiliation       & \begin{tabular}[c]{@{}l@{}}The relational position of an entity, reflecting both its dynamic attitude \\ (supportive, adversarial,   competitive) or its enduring association (group membership, loyalty).\end{tabular} & \begin{tabular}[c]{@{}l@{}}Enemy/Antagonist\\      Ally/Companion\\      Rival/Competitor\end{tabular}                                    \\ \midrule
Stranger                  & Stranger                                                                                                                                                                                                                & \begin{tabular}[c]{@{}l@{}}Stranger \end{tabular}                                       \\ \bottomrule
\end{tabular}}
\caption{Taxonomy of Annotated Relationship Categories and Subcategories.}
\label{tab:relationship}
\end{table*}

\begin{table*}[h!]
\resizebox{\textwidth}{!}{
\begin{tabular}{@{}|l|l|l|l|l|@{}}
\toprule
\textbf{Category}                      & \textbf{Description}                                                                                                                              & \textbf{Subcategories}    & \textbf{Traits}          & \textbf{Description}                                                                                       \\ \midrule
\multirow{18}{*}{Moral \&   Ethical}   & \multirow{18}{*}{\begin{tabular}[c]{@{}l@{}}Describes a character's fundamental sense of right \\ and wrong and how they act on it.\end{tabular}} & \multirow{6}{*}{positive} & Altruistic               & Selflessly devoted to the welfare of others.                                                               \\ \cmidrule(l){4-5} 
                                       &                                                                                                                                                   &                           & Compassionate/Empathetic & Feels and understands the suffering of others.                                                             \\ \cmidrule(l){4-5} 
                                       &                                                                                                                                                   &                           & Honorable/Principled     & Adheres to a strict   moral code.                                                                          \\ \cmidrule(l){4-5} 
                                       &                                                                                                                                                   &                           & Just/Fair-minded         & Strives to treat everyone equitably.                                                                       \\ \cmidrule(l){4-5} 
                                       &                                                                                                                                                   &                           & Loyal                    & Faithful to a person, cause, or group.                                                                     \\ \cmidrule(l){4-5} 
                                       &                                                                                                                                                   &                           & Merciful/Forgiving       & Shows compassion to those they have power over.                                                            \\ \cmidrule(l){3-5} 
                                       &                                                                                                                                                   & \multirow{6}{*}{negative} & Selfish/Egotistical      & Concerned primarily with their own interests.                                                              \\ \cmidrule(l){4-5} 
                                       &                                                                                                                                                   &                           & Cruel/Sadistic           & Enjoys inflicting pain or suffering.                                                                       \\ \cmidrule(l){4-5} 
                                       &                                                                                                                                                   &                           & Unscrupulous/Corrupt     & Lacking moral principles; willing to cheat or lie.                                                         \\ \cmidrule(l){4-5} 
                                       &                                                                                                                                                   &                           & Prejudiced/Biased        & Holds preconceived, often unfair, judgments.                                                               \\ \cmidrule(l){4-5} 
                                       &                                                                                                                                                   &                           & Treacherous/Deceitful    & Betrays trust; disloyal.                                                                                   \\ \cmidrule(l){4-5} 
                                       &                                                                                                                                                   &                           & Vindictive/Spiteful      & Seeks revenge; holds grudges.                                                                              \\ \cmidrule(l){3-5} 
                                       &                                                                                                                                                   & \multirow{6}{*}{neutral:} & Pragmatic                & Guided by practical results rather than ideals.                                                            \\ \cmidrule(l){4-5} 
                                       &                                                                                                                                                   &                           & Utilitarian              & Believes actions are right if they benefit the majority.                                                   \\ \cmidrule(l){4-5} 
                                       &                                                                                                                                                   &                           & Obedient/Dutiful         & Follows orders or a prescribed code without question.                                                      \\ \cmidrule(l){4-5} 
                                       &                                                                                                                                                   &                           & Detached                 & Able to separate from moral or emotional entanglement.                                                     \\ \cmidrule(l){4-5} 
                                       &                                                                                                                                                   &                           & Opportunistic            & \begin{tabular}[c]{@{}l@{}}Adapts actions to seize opportunities, \\ regardless of principle.\end{tabular} \\ \cmidrule(l){4-5} 
                                       &                                                                                                                                                   &                           & Survivalist              & Prioritizes self-preservation above all else.                                                              \\ \midrule
\multirow{15}{*}{Social   Disposition} & \multirow{15}{*}{\begin{tabular}[c]{@{}l@{}}Describes how a  character behaves in social settings \\ and builds relationships.\end{tabular}}      & \multirow{5}{*}{positive} & Affable/Amiable          & Friendly, good-natured, and easy to talk to.                                                               \\ \cmidrule(l){4-5} 
                                       &                                                                                                                                                   &                           & Charismatic/Charming     & Has a magnetic quality that inspires devotion in others.                                                   \\ \cmidrule(l){4-5} 
                                       &                                                                                                                                                   &                           & Diplomatic/Tactful       & Skilled in dealing with people sensitively and effectively.                                                \\ \cmidrule(l){4-5} 
                                       &                                                                                                                                                   &                           & Gregarious               & Fond of company; sociable.                                                                                 \\ \cmidrule(l){4-5} 
                                       &                                                                                                                                                   &                           & Trusting                 & Believes in the reliability and good intentions of others.                                                 \\ \cmidrule(l){3-5} 
                                       &                                                                                                                                                   & \multirow{5}{*}{negative} & Misanthropic             & Dislikes and distrusts humankind.                                                                          \\ \cmidrule(l){4-5} 
                                       &                                                                                                                                                   &                           & Abrasive/Confrontational & Grating and harsh in their social interactions.                                                            \\ \cmidrule(l){4-5} 
                                       &                                                                                                                                                   &                           & Manipulative             & Controls or influences others cleverly and unscrupulously.                                                 \\ \cmidrule(l){4-5} 
                                       &                                                                                                                                                   &                           & Aloof/Standoffish        & Distant and cool in manner.                                                                                \\ \cmidrule(l){4-5} 
                                       &                                                                                                                                                   &                           & Gullible/Naïve           & Easily persuaded to believe something; too trusting.                                                       \\ \cmidrule(l){3-5} 
                                       &                                                                                                                                                   & \multirow{5}{*}{neutral}  & Introverted              & Energized by solitude; may be quiet or reserved.                                                           \\ \cmidrule(l){4-5} 
                                       &                                                                                                                                                   &                           & Extroverted              & Energized by social interaction; often outgoing.                                                           \\ \cmidrule(l){4-5} 
                                       &                                                                                                                                                   &                           & Reserved/Private         & Tends to keep their thoughts and feelings to themselves.                                                   \\ \cmidrule(l){4-5} 
                                       &                                                                                                                                                   &                           & Unassuming/Modest        & Not pretentious or arrogant.                                                                               \\ \cmidrule(l){4-5} 
                                       &                                                                                                                                                   &                           & Formal                   & Adheres strictly to convention and etiquette.                                                              \\ \bottomrule
\end{tabular}}
\caption{Taxonomy of Annotated Personality Categories and Subcategories (1).}
\label{tab:instinct_traits_1}
\end{table*}

\begin{table*}[h!]
\resizebox{\textwidth}{!}{
\begin{tabular}{@{}|l|l|l|l|l|@{}}
\toprule
\textbf{Category}                                                                                        & \textbf{Description}                                                                                                                        & \textbf{Subcategories}    & \textbf{Traits}             & \textbf{Description}                                                                 \\ \midrule
\multicolumn{1}{|c|}{\multirow{15}{*}{\begin{tabular}[c]{@{}c@{}}Approach to\\  Adversity\end{tabular}}} & \multirow{15}{*}{\begin{tabular}[c]{@{}l@{}}Focuses on a character's courage, determination, \\ and reaction to adversity.\end{tabular}}    & \multirow{5}{*}{positive} & Brave/Courageous            & Faces danger, pain, or   difficulty with fortitude.                                  \\ \cmidrule(l){4-5} 
\multicolumn{1}{|c|}{}                                                                                   &                                                                                                                                             &                           & Resilient                   & Able to recover quickly from difficulties; tough.                                    \\ \cmidrule(l){4-5} 
\multicolumn{1}{|c|}{}                                                                                   &                                                                                                                                             &                           & Determined/Persistent       & Continues firmly in a course of action despite difficulty.                           \\ \cmidrule(l){4-5} 
\multicolumn{1}{|c|}{}                                                                                   &                                                                                                                                             &                           & Proactive/Ambitious         & Acts in anticipation of future problems or needs; driven to   succeed.               \\ \cmidrule(l){4-5} 
\multicolumn{1}{|c|}{}                                                                                   &                                                                                                                                             &                           & Audacious/Daring            & Willing to take surprising or bold risks.                                            \\ \cmidrule(l){3-5} 
\multicolumn{1}{|c|}{}                                                                                   &                                                                                                                                             & \multirow{5}{*}{negative} & Timid/Cowardly              & Lacks courage; is easily frightened.                                                 \\ \cmidrule(l){4-5} 
\multicolumn{1}{|c|}{}                                                                                   &                                                                                                                                             &                           & Fatalistic                  & Believes all events are predetermined and therefore   inevitable.                    \\ \cmidrule(l){4-5} 
\multicolumn{1}{|c|}{}                                                                                   &                                                                                                                                             &                           & Apathetic/Complacent        & Lacks interest, enthusiasm, or concern.                                              \\ \cmidrule(l){4-5} 
\multicolumn{1}{|c|}{}                                                                                   &                                                                                                                                             &                           & Indecisive/Vacillating      & Unable to make decisions quickly and effectively.                                    \\ \cmidrule(l){4-5} 
\multicolumn{1}{|c|}{}                                                                                   &                                                                                                                                             &                           & Reckless/Impulsive          & Acts without thinking about the consequences.                                        \\ \cmidrule(l){3-5} 
\multicolumn{1}{|c|}{}                                                                                   &                                                                                                                                             & \multirow{5}{*}{neutral}  & Cautious                    & Careful to avoid potential problems or dangers.                                      \\ \cmidrule(l){4-5} 
\multicolumn{1}{|c|}{}                                                                                   &                                                                                                                                             &                           & Stoic                       & Endures pain or hardship without showing feelings or   complaining.                  \\ \cmidrule(l){4-5} 
\multicolumn{1}{|c|}{}                                                                                   &                                                                                                                                             &                           & Spontaneous                 & Acts as a result of a sudden inner impulse.                                          \\ \cmidrule(l){4-5} 
\multicolumn{1}{|c|}{}                                                                                   &                                                                                                                                             &                           & Methodical                  & Does things in an orderly, systematic way.                                           \\ \cmidrule(l){4-5} 
\multicolumn{1}{|c|}{}                                                                                   &                                                                                                                                             &                           & Passive                     & Accepts or allows what happens without active response.                              \\ \midrule
\multirow{15}{*}{Self-Perception}                                                                        & \multirow{15}{*}{\begin{tabular}[c]{@{}l@{}}Relates to a character's view of themselves \\ and their typical emotional state.\end{tabular}} & \multirow{5}{*}{positive} & Confident/Self-assured      & Trusts in their own abilities and judgment.                                          \\ \cmidrule(l){4-5} 
                                                                                                         &                                                                                                                                             &                           & Humble/Modest               & Has a low view of their own importance.                                              \\ \cmidrule(l){4-5} 
                                                                                                         &                                                                                                                                             &                           & Optimistic/Sanguine         & Hopeful and confident about the future.                                              \\ \cmidrule(l){4-5} 
                                                                                                         &                                                                                                                                             &                           & Serene/Placid               & Calm, peaceful, and untroubled.                                                      \\ \cmidrule(l){4-5} 
                                                                                                         &                                                                                                                                             &                           & Disciplined/Self-controlled & Able to control their feelings and overcome weaknesses.                              \\ \cmidrule(l){3-5} 
                                                                                                         &                                                                                                                                             & \multirow{5}{*}{negative} & Arrogant/Haughty            & Has an exaggerated sense of their own importance.                                    \\ \cmidrule(l){4-5} 
                                                                                                         &                                                                                                                                             &                           & Insecure/Self-doubting      & Lacks confidence in themselves.                                                      \\ \cmidrule(l){4-5} 
                                                                                                         &                                                                                                                                             &                           & Pessimistic/Cynical         & Tends to see the worst in things or believe people are   motivated by self-interest. \\ \cmidrule(l){4-5} 
                                                                                                         &                                                                                                                                             &                           & Irascible/Volatile          & Easily angered; prone to outbursts.                                                  \\ \cmidrule(l){4-5} 
                                                                                                         &                                                                                                                                             &                           & Vain/Narcissistic           & Has an excessive interest in or admiration of themselves.                            \\ \cmidrule(l){3-5} 
                                                                                                         &                                                                                                                                             & \multirow{5}{*}{neutral}  & Introspective               & Examines their own thoughts and feelings.                                            \\ \cmidrule(l){4-5} 
                                                                                                         &                                                                                                                                             &                           & Proud                       & Has a deep pleasure or satisfaction from one's own   achievements.                   \\ \cmidrule(l){4-5} 
                                                                                                         &                                                                                                                                             &                           & Ambitious                   & Has a strong desire for success, power, or fame.                                     \\ \cmidrule(l){4-5} 
                                                                                                         &                                                                                                                                             &                           & Earnest                     & Shows sincere and intense conviction.                                                \\ \cmidrule(l){4-5} 
                                                                                                         &                                                                                                                                             &                           & Melancholic                 & Prone to a feeling of pensive sadness, typically with no   obvious cause.            \\ \bottomrule
\end{tabular}}
\caption{Taxonomy of Annotated Personality Categories and Subcategories (2).}
\label{tab:instinct_traits_2}
\end{table*}

%% file: main.bbl
\begin{thebibliography}{10}
\providecommand{\url}[1]{#1}
\csname url@samestyle\endcsname
\providecommand{\newblock}{\relax}
\providecommand{\bibinfo}[2]{#2}
\providecommand{\BIBentrySTDinterwordspacing}{\spaceskip=0pt\relax}
\providecommand{\BIBentryALTinterwordstretchfactor}{4}
\providecommand{\BIBentryALTinterwordspacing}{\spaceskip=\fontdimen2\font plus
\BIBentryALTinterwordstretchfactor\fontdimen3\font minus \fontdimen4\font\relax}
\providecommand{\BIBforeignlanguage}[2]{{%
\expandafter\ifx\csname l@#1\endcsname\relax
\typeout{** WARNING: IEEEtran.bst: No hyphenation pattern has been}%
\typeout{** loaded for the language `#1'. Using the pattern for}%
\typeout{** the default language instead.}%
\else
\language=\csname l@#1\endcsname
\fi
#2}}
\providecommand{\BIBdecl}{\relax}
\BIBdecl

\bibitem{cai}
\BIBentryALTinterwordspacing
Character.AI, ``Character AI,'' 2023. [Online]. Available: \url{https://c.ai}
\BIBentrySTDinterwordspacing

\bibitem{cai_contrary}
\BIBentryALTinterwordspacing
ContraryResearch, ``Character AI,'' 2024. [Online]. Available: \url{https://research.contrary.com/company/character-ai}
\BIBentrySTDinterwordspacing

\bibitem{janitorAI}
\BIBentryALTinterwordspacing
JanitorAI, ``Janitor AI,'' 2023. [Online]. Available: \url{https://janitorai.com}
\BIBentrySTDinterwordspacing

\bibitem{Student5541694_2024}
J.~Li and S.~5541694, ``The Formation and Maintenance of AI Intimacy: A Study on Character.AI,'' \emph{University of Warwick}, August 2024, mA Digital Media and Culture Dissertation, Centre for Interdisciplinary Methodologies.

\bibitem{liu2024chatbotcompanionshipmixedmethodsstudy}
\BIBentryALTinterwordspacing
A.~R. Liu, P.~Pataranutaporn, and P.~Maes, ``Chatbot Companionship: A Mixed-Methods Study of Companion Chatbot Usage Patterns and Their Relationship to Loneliness in Active Users,'' 2024. [Online]. Available: \url{https://arxiv.org/abs/2410.21596}
\BIBentrySTDinterwordspacing

\bibitem{Guingrich_2025}
\BIBentryALTinterwordspacing
R.~E. Guingrich and M.~S.~A. Graziano, \emph{Chatbots as Social Companions: How People Perceive Consciousness, Human Likeness, and Social Health Benefits in Machines}.\hskip 1em plus 0.5em minus 0.4em\relax Oxford University PressOxford, Mar. 2025. [Online]. Available: \url{http://dx.doi.org/10.1093/9780198945215.003.0011}
\BIBentrySTDinterwordspacing

\bibitem{chandra2024livedexperienceinsightunpacking}
\BIBentryALTinterwordspacing
M.~Chandra, S.~Naik, D.~Ford, E.~Okoli, M.~D. Choudhury, M.~Ershadi, G.~Ramos, J.~Hernandez, A.~Bhattacharjee, S.~Warreth, and J.~Suh, ``From Lived Experience to Insight: Unpacking the Psychological Risks of Using AI Conversational Agents,'' 2024. [Online]. Available: \url{https://arxiv.org/abs/2412.07951}
\BIBentrySTDinterwordspacing

\bibitem{yu2024exploringparentchildperceptionssafety}
\BIBentryALTinterwordspacing
Y.~Yu, T.~Sharma, M.~Hu, J.~Wang, and Y.~Wang, ``Exploring Parent-Child Perceptions on Safety in Generative AI: Concerns, Mitigation Strategies, and Design Implications,'' 2024. [Online]. Available: \url{https://arxiv.org/abs/2406.10461}
\BIBentrySTDinterwordspacing

\bibitem{DeFreitas2025}
\BIBentryALTinterwordspacing
J.~De~Freitas and I.~Cohen, ``Unregulated emotional risks of AI wellness apps,'' \emph{Nature Machine Intelligence}, vol.~7, pp. 813--815, June 2025. [Online]. Available: \url{https://doi.org/10.1038/s42256-025-01051-5}
\BIBentrySTDinterwordspacing

\bibitem{zhang2025realherexploringyoung}
\BIBentryALTinterwordspacing
S.~Zhang and S.~Li, ``The Real Her? Exploring Whether Young Adults Accept Human-AI Love,'' 2025. [Online]. Available: \url{https://arxiv.org/abs/2503.03067}
\BIBentrySTDinterwordspacing

\bibitem{10.1145/3687022}
\BIBentryALTinterwordspacing
H.~Chin, A.~Zhunis, and M.~Cha, ``Behaviors and Perceptions of Human-Chatbot Interactions Based on Top Active Users of a Commercial Social Chatbot,'' \emph{Proc. ACM Hum.-Comput. Interact.}, vol.~8, no. CSCW2, Nov. 2024. [Online]. Available: \url{https://doi.org/10.1145/3687022}
\BIBentrySTDinterwordspacing

\bibitem{yu2025understandinggenerativeairisks}
\BIBentryALTinterwordspacing
Y.~Yu, Y.~Liu, J.~Zhang, Y.~Huang, and Y.~Wang, ``Understanding Generative AI Risks for Youth: A Taxonomy Based on Empirical Data,'' 2025. [Online]. Available: \url{https://arxiv.org/abs/2502.16383}
\BIBentrySTDinterwordspacing

\bibitem{qiu2025emoagentassessingsafeguardinghumanai}
\BIBentryALTinterwordspacing
J.~Qiu, Y.~He, X.~Juan, Y.~Wang, Y.~Liu, Z.~Yao, Y.~Wu, X.~Jiang, L.~Yang, and M.~Wang, ``EmoAgent: Assessing and Safeguarding Human-AI Interaction for Mental Health Safety,'' 2025. [Online]. Available: \url{https://arxiv.org/abs/2504.09689}
\BIBentrySTDinterwordspacing

\bibitem{qut2023chatbots}
\BIBentryALTinterwordspacing
H.~Fraser, ``Deaths Linked to Chatbots Show We Must Urgently Revisit What Counts as High-Risk AI,'' 2023. [Online]. Available: \url{https://www.qut.edu.au/news/realfocus/deaths-linked-to-chatbots-show-we-must-urgently-revisit-what-counts-as-high-risk-ai}
\BIBentrySTDinterwordspacing

\bibitem{ji2025enhancingpersonaconsistencyllms}
\BIBentryALTinterwordspacing
K.~Ji, Y.~Lian, L.~Li, J.~Gao, W.~Li, and B.~Dai, ``Enhancing Persona Consistency for LLMs' Role-Playing using Persona-Aware Contrastive Learning,'' 2025. [Online]. Available: \url{https://arxiv.org/abs/2503.17662}
\BIBentrySTDinterwordspacing

\bibitem{tseng2024talespersonallmssurvey}
\BIBentryALTinterwordspacing
Y.-M. Tseng, Y.-C. Huang, T.-Y. Hsiao, W.-L. Chen, C.-W. Huang, Y.~Meng, and Y.-N. Chen, ``Two Tales of Persona in LLMs: A Survey of Role-Playing and Personalization,'' 2024. [Online]. Available: \url{https://arxiv.org/abs/2406.01171}
\BIBentrySTDinterwordspacing

\bibitem{chen2024personapersonalizationsurveyroleplaying}
\BIBentryALTinterwordspacing
J.~Chen, X.~Wang, R.~Xu, S.~Yuan, Y.~Zhang, W.~Shi, J.~Xie, S.~Li, R.~Yang, T.~Zhu, A.~Chen, N.~Li, L.~Chen, C.~Hu, S.~Wu, S.~Ren, Z.~Fu, and Y.~Xiao, ``From Persona to Personalization: A Survey on Role-Playing Language Agents,'' 2024. [Online]. Available: \url{https://arxiv.org/abs/2404.18231}
\BIBentrySTDinterwordspacing

\bibitem{chen2025oscarsaitheatersurvey}
\BIBentryALTinterwordspacing
N.~Chen, Y.~Wang, Y.~Deng, and J.~Li, ``The Oscars of AI Theater: A Survey on Role-Playing with Language Models,'' 2025. [Online]. Available: \url{https://arxiv.org/abs/2407.11484}
\BIBentrySTDinterwordspacing

\bibitem{betley2025emergentmisalignmentnarrowfinetuning}
\BIBentryALTinterwordspacing
J.~Betley, D.~Tan, N.~Warncke, A.~Sztyber-Betley, X.~Bao, M.~Soto, N.~Labenz, and O.~Evans, ``Emergent Misalignment: Narrow finetuning can produce broadly misaligned LLMs,'' 2025. [Online]. Available: \url{https://arxiv.org/abs/2502.17424}
\BIBentrySTDinterwordspacing

\bibitem{qi2023finetuningalignedlanguagemodels}
\BIBentryALTinterwordspacing
X.~Qi, Y.~Zeng, T.~Xie, P.-Y. Chen, R.~Jia, P.~Mittal, and P.~Henderson, ``Fine-tuning Aligned Language Models Compromises Safety, Even When Users Do Not Intend To!'' 2023. [Online]. Available: \url{https://arxiv.org/abs/2310.03693}
\BIBentrySTDinterwordspacing

\bibitem{jin2024jailbreakzoosurveylandscapeshorizons}
\BIBentryALTinterwordspacing
H.~Jin, L.~Hu, X.~Li, P.~Zhang, C.~Chen, J.~Zhuang, and H.~Wang, ``JailbreakZoo: Survey, Landscapes, and Horizons in Jailbreaking Large Language and Vision-Language Models,'' 2024. [Online]. Available: \url{https://arxiv.org/abs/2407.01599}
\BIBentrySTDinterwordspacing

\bibitem{yi2024jailbreakattacksdefenseslarge}
\BIBentryALTinterwordspacing
S.~Yi, Y.~Liu, Z.~Sun, T.~Cong, X.~He, J.~Song, K.~Xu, and Q.~Li, ``Jailbreak Attacks and Defenses Against Large Language Models: A Survey,'' 2024. [Online]. Available: \url{https://arxiv.org/abs/2407.04295}
\BIBentrySTDinterwordspacing

\bibitem{li-etal-2024-salad}
\BIBentryALTinterwordspacing
L.~Li, B.~Dong, R.~Wang, X.~Hu, W.~Zuo, D.~Lin, Y.~Qiao, and J.~Shao, ``{SALAD}-Bench: A Hierarchical and Comprehensive Safety Benchmark for Large Language Models,'' in \emph{Findings of the Association for Computational Linguistics: ACL 2024}, L.-W. Ku, A.~Martins, and V.~Srikumar, Eds.\hskip 1em plus 0.5em minus 0.4em\relax Bangkok, Thailand: Association for Computational Linguistics, Aug. 2024, pp. 3923--3954. [Online]. Available: \url{https://aclanthology.org/2024.findings-acl.235/}
\BIBentrySTDinterwordspacing

\bibitem{Röttger_Pernisi_Vidgen_Hovy_2025}
\BIBentryALTinterwordspacing
P.~Röttger, F.~Pernisi, B.~Vidgen, and D.~Hovy, ``SafetyPrompts: A Systematic Review of Open Datasets for Evaluating and Improving Large Language Model Safety,'' \emph{Proceedings of the AAAI Conference on Artificial Intelligence}, vol.~39, no.~26, pp. 27\,617--27\,627, Apr. 2025. [Online]. Available: \url{https://ojs.aaai.org/index.php/AAAI/article/view/34975}
\BIBentrySTDinterwordspacing

\bibitem{liu2025scalesjustitiacomprehensivesurvey}
\BIBentryALTinterwordspacing
S.~Liu, C.~Li, J.~Qiu, X.~Zhang, F.~Huang, L.~Zhang, Y.~Hei, and P.~S. Yu, ``The Scales of Justitia: A Comprehensive Survey on Safety Evaluation of LLMs,'' 2025. [Online]. Available: \url{https://arxiv.org/abs/2506.11094}
\BIBentrySTDinterwordspacing

\bibitem{bai2022traininghelpfulharmlessassistant}
\BIBentryALTinterwordspacing
Y.~Bai, A.~Jones, K.~Ndousse, A.~Askell, A.~Chen, N.~DasSarma, D.~Drain, S.~Fort, D.~Ganguli, T.~Henighan, N.~Joseph, S.~Kadavath, J.~Kernion, T.~Conerly, S.~El-Showk, N.~Elhage, Z.~Hatfield-Dodds, D.~Hernandez, T.~Hume, S.~Johnston, S.~Kravec, L.~Lovitt, N.~Nanda, C.~Olsson, D.~Amodei, T.~Brown, J.~Clark, S.~McCandlish, C.~Olah, B.~Mann, and J.~Kaplan, ``Training a Helpful and Harmless Assistant with Reinforcement Learning from Human Feedback,'' 2022. [Online]. Available: \url{https://arxiv.org/abs/2204.05862}
\BIBentrySTDinterwordspacing

\bibitem{ganguli2022redteaminglanguagemodels}
\BIBentryALTinterwordspacing
D.~Ganguli, L.~Lovitt, J.~Kernion, A.~Askell, Y.~Bai, S.~Kadavath, B.~Mann, E.~Perez, N.~Schiefer, K.~Ndousse, A.~Jones, S.~Bowman, A.~Chen, T.~Conerly, N.~DasSarma, D.~Drain, N.~Elhage, S.~El-Showk, S.~Fort, Z.~Hatfield-Dodds, T.~Henighan, D.~Hernandez, T.~Hume, J.~Jacobson, S.~Johnston, S.~Kravec, C.~Olsson, S.~Ringer, E.~Tran-Johnson, D.~Amodei, T.~Brown, N.~Joseph, S.~McCandlish, C.~Olah, J.~Kaplan, and J.~Clark, ``Red Teaming Language Models to Reduce Harms: Methods, Scaling Behaviors, and Lessons Learned,'' 2022. [Online]. Available: \url{https://arxiv.org/abs/2209.07858}
\BIBentrySTDinterwordspacing

\bibitem{yu2024gptfuzzerredteaminglarge}
\BIBentryALTinterwordspacing
J.~Yu, X.~Lin, Z.~Yu, and X.~Xing, ``GPTFUZZER: Red Teaming Large Language Models with Auto-Generated Jailbreak Prompts,'' 2024. [Online]. Available: \url{https://arxiv.org/abs/2309.10253}
\BIBentrySTDinterwordspacing

\bibitem{10.1145/3658644.3670388}
\BIBentryALTinterwordspacing
X.~Shen, Z.~Chen, M.~Backes, Y.~Shen, and Y.~Zhang, ````Do Anything Now'': Characterizing and Evaluating In-The-Wild Jailbreak Prompts on Large Language Models,'' in \emph{Proceedings of the 2024 on ACM SIGSAC Conference on Computer and Communications Security}, ser. CCS '24.\hskip 1em plus 0.5em minus 0.4em\relax New York, NY, USA: Association for Computing Machinery, 2024, p. 1671–1685. [Online]. Available: \url{https://doi.org/10.1145/3658644.3670388}
\BIBentrySTDinterwordspacing

\bibitem{deng2024multilingual}
\BIBentryALTinterwordspacing
Y.~Deng, W.~Zhang, S.~J. Pan, and L.~Bing, ``Multilingual Jailbreak Challenges in Large Language Models,'' in \emph{The Twelfth International Conference on Learning Representations}, 2024. [Online]. Available: \url{https://openreview.net/forum?id=vESNKdEMGp}
\BIBentrySTDinterwordspacing

\bibitem{wang-etal-2024-answer}
\BIBentryALTinterwordspacing
Y.~Wang, H.~Li, X.~Han, P.~Nakov, and T.~Baldwin, ``Do-Not-Answer: Evaluating Safeguards in {LLM}s,'' in \emph{Findings of the Association for Computational Linguistics: EACL 2024}, Y.~Graham and M.~Purver, Eds.\hskip 1em plus 0.5em minus 0.4em\relax St. Julian{'}s, Malta: Association for Computational Linguistics, Mar. 2024, pp. 896--911. [Online]. Available: \url{https://aclanthology.org/2024.findings-eacl.61/}
\BIBentrySTDinterwordspacing

\bibitem{lin-etal-2023-toxicchat}
\BIBentryALTinterwordspacing
Z.~Lin, Z.~Wang, Y.~Tong, Y.~Wang, Y.~Guo, Y.~Wang, and J.~Shang, ``{T}oxic{C}hat: Unveiling Hidden Challenges of Toxicity Detection in Real-World User-{AI} Conversation,'' in \emph{Findings of the Association for Computational Linguistics: EMNLP 2023}, H.~Bouamor, J.~Pino, and K.~Bali, Eds.\hskip 1em plus 0.5em minus 0.4em\relax Singapore: Association for Computational Linguistics, Dec. 2023, pp. 4694--4702. [Online]. Available: \url{https://aclanthology.org/2023.findings-emnlp.311/}
\BIBentrySTDinterwordspacing

\bibitem{zou2023universaltransferableadversarialattacks}
\BIBentryALTinterwordspacing
A.~Zou, Z.~Wang, N.~Carlini, M.~Nasr, J.~Z. Kolter, and M.~Fredrikson, ``Universal and Transferable Adversarial Attacks on Aligned Language Models,'' 2023. [Online]. Available: \url{https://arxiv.org/abs/2307.15043}
\BIBentrySTDinterwordspacing

\bibitem{cao2025safedialbenchfinegrainedsafetybenchmark}
\BIBentryALTinterwordspacing
H.~Cao, Y.~Wang, S.~Jing, Z.~Peng, Z.~Bai, Z.~Cao, M.~Fang, F.~Feng, B.~Wang, J.~Liu, T.~Yang, J.~Huo, Y.~Gao, F.~Meng, X.~Yang, C.~Deng, and J.~Feng, ``SafeDialBench: A Fine-Grained Safety Benchmark for Large Language Models in Multi-Turn Dialogues with Diverse Jailbreak Attacks,'' 2025. [Online]. Available: \url{https://arxiv.org/abs/2502.11090}
\BIBentrySTDinterwordspacing

\bibitem{md_judge}
OpenSafetyLab, ``MD-Judge-v0\_2-internlm2\_7b,'' \url{https://huggingface.co/OpenSafetyLab/MD-Judge-v0_2-internlm2_7b}, 2024.

\bibitem{li2024benchmarkingroleplaying}
X.~Li, Z.~Chen, J.~M. Zhang, Y.~Lou, T.~Li, W.~Sun, Y.~Liu, and X.~Liu, ``Benchmarking Bias in Large Language Models during Role-Playing,'' \emph{arXiv preprint arXiv:2411.00585}, 2024.

\bibitem{lee2025promptingfailsswayinertia}
\BIBentryALTinterwordspacing
B.~W. Lee, Y.~Lee, and H.~Cho, ``When Prompting Fails to Sway: Inertia in Moral and Value Judgments of Large Language Models,'' 2025. [Online]. Available: \url{https://arxiv.org/abs/2408.09049}
\BIBentrySTDinterwordspacing

\bibitem{kim2025exploringpersonadependentllmalignment}
\BIBentryALTinterwordspacing
J.~Kim, J.~Kwon, L.~F. Vecchietti, A.~Oh, and M.~Cha, ``Exploring Persona-dependent LLM Alignment for the Moral Machine Experiment,'' 2025. [Online]. Available: \url{https://arxiv.org/abs/2504.10886}
\BIBentrySTDinterwordspacing

\bibitem{Forster1927Aspects}
E.~M. Forster, \emph{Aspects of the Novel}.\hskip 1em plus 0.5em minus 0.4em\relax London: Edward Arnold, 1927.

\bibitem{Schmidt2001Master}
V.~L. Schmidt, \emph{45 Master Characters: Mythic Models for Creating Original Characters}.\hskip 1em plus 0.5em minus 0.4em\relax Cincinnati, OH: Writer's Digest Books, 2001.

\bibitem{Tally2012Spatiality}
R.~T.~J. Tally, \emph{Spatiality}, ser. The New Critical Idiom.\hskip 1em plus 0.5em minus 0.4em\relax London: Routledge, 2012.

\bibitem{Carty2021Inside}
S.~Carty, \emph{Inside Fictional Minds: Tips from Psychology for Creating Characters}.\hskip 1em plus 0.5em minus 0.4em\relax UK: Ad Hoc Fiction, 2021.

\bibitem{Tally2017Routledge}
R.~T.~J. Tally, ``The Routledge Handbook of Literature and Space,'' London, 2017.

\bibitem{nonparametric_statistics}
S.~Siegel and N.~J. Castellan~Jr., \emph{Nonparametric Statistics for the Behavioral Sciences}, 2nd~ed.\hskip 1em plus 0.5em minus 0.4em\relax New York: McGraw-Hill, 1988.

\bibitem{chi-square-goodness-of-fit}
\BIBentryALTinterwordspacing
S.~K. Portal, ``Chi-Square Goodness of Fit Test,'' 2025. [Online]. Available: \url{https://www.jmp.com/en/statistics-knowledge-portal/chi-square-test/chi-square-goodness-of-fit-test}
\BIBentrySTDinterwordspacing

\bibitem{kruskal1952use}
W.~H. Kruskal and W.~A. Wallis, ``Use of ranks in one-criterion variance analysis,'' \emph{Journal of the American statistical association}, vol.~47, no. 260, pp. 583--621, 1952.

\bibitem{Montgomery2018Applied}
D.~C. Montgomery and G.~C. Runger, \emph{Applied Statistics and Probability for Engineers}, 7th~ed.\hskip 1em plus 0.5em minus 0.4em\relax Hoboken, NJ: John Wiley \& Sons, 2018.

\bibitem{breiman_permutation_2001}
L.~Breiman, ``Random forests,'' \emph{Machine Learning}, vol.~45, no.~1, pp. 5--32, 2001.

\bibitem{10.1177/21677026231186625}
\BIBentryALTinterwordspacing
V.~M.~E. Bridgland, P.~J. Jones, and B.~W. Bellet, ``A Meta-Analysis of the Efficacy of Trigger Warnings, Content Warnings, and Content Notes,'' \emph{Clinical Psychological Science}, vol.~12, no.~4, pp. 751--771, 2024. [Online]. Available: \url{https://doi.org/10.1177/21677026231186625}
\BIBentrySTDinterwordspacing

\bibitem{li2023chatharuhi}
C.~Li, Z.~Leng, C.~Yan, J.~Shen, H.~Wang, W.~Mi, Y.~Fei, X.~Feng, S.~Yan, H.~Wang \emph{et~al.}, ``Chatharuhi: Reviving anime character in reality via large language model,'' \emph{arXiv preprint arXiv:2308.09597}, 2023.

\bibitem{schick2023toolformer}
T.~Schick, J.~Dwivedi-Yu, R.~Dess{\`\i}, R.~Raileanu, M.~Lomeli, E.~Hambro, L.~Zettlemoyer, N.~Cancedda, and T.~Scialom, ``Toolformer: Language models can teach themselves to use tools,'' \emph{Advances in Neural Information Processing Systems}, vol.~36, pp. 68\,539--68\,551, 2023.

\bibitem{li2023camel}
G.~Li, H.~Hammoud, H.~Itani, D.~Khizbullin, and B.~Ghanem, ``Camel: Communicative agents for" mind" exploration of large language model society,'' \emph{Advances in Neural Information Processing Systems}, vol.~36, pp. 51\,991--52\,008, 2023.

\bibitem{chen2023autoagents}
G.~Chen, S.~Dong, Y.~Shu, G.~Zhang, J.~Sesay, B.~F. Karlsson, J.~Fu, and Y.~Shi, ``Autoagents: A framework for automatic agent generation,'' \emph{arXiv preprint arXiv:2309.17288}, 2023.

\bibitem{zhao2025beware}
W.~Zhao, Y.~Hu, Y.~Deng, J.~Guo, X.~Sui, X.~Han, A.~Zhang, Y.~Zhao, B.~Qin, T.-S. Chua \emph{et~al.}, ``Beware of your po! measuring and mitigating ai safety risks in role-play fine-tuning of llms,'' \emph{arXiv preprint arXiv:2502.20968}, 2025.

\bibitem{kamruzzaman2024woman}
M.~Kamruzzaman, H.~Nguyen, N.~Hassan, and G.~L. Kim, ``" A Woman is More Culturally Knowledgeable than A Man?": The Effect of Personas on Cultural Norm Interpretation in LLMs,'' \emph{arXiv preprint arXiv:2409.11636}, 2024.

\bibitem{zhao2024bias}
J.~Zhao, Z.~Qian, L.~Cao, Y.~Wang, and Y.~Ding, ``Bias and toxicity in role-play reasoning,'' \emph{arXiv e-prints}, pp. arXiv--2409, 2024.

\bibitem{zhang2024better}
J.~Zhang, D.~Liu, C.~Qian, Z.~Gan, Y.~Liu, Y.~Qiao, and J.~Shao, ``The better angels of machine personality: How personality relates to llm safety,'' \emph{arXiv preprint arXiv:2407.12344}, 2024.

\bibitem{naous2023beer}
T.~Naous, M.~J. Ryan, A.~Ritter, and W.~Xu, ``Having beer after prayer? measuring cultural bias in large language models,'' \emph{arXiv preprint arXiv:2305.14456}, 2023.

\bibitem{zhang2025dark}
R.~Zhang, H.~Li, H.~Meng, J.~Zhan, H.~Gan, and Y.-C. Lee, ``The dark side of ai companionship: A taxonomy of harmful algorithmic behaviors in human-ai relationships,'' in \emph{Proceedings of the 2025 CHI Conference on Human Factors in Computing Systems}, 2025, pp. 1--17.

\bibitem{chu2025illusions}
M.~D. Chu, P.~Gerard, K.~Pawar, C.~Bickham, and K.~Lerman, ``Illusions of intimacy: Emotional attachment and emerging psychological risks in human-ai relationships,'' \emph{arXiv preprint arXiv:2505.11649}, 2025.

\bibitem{zhang2025rise}
Y.~Zhang, D.~Zhao, J.~T. Hancock, R.~Kraut, and D.~Yang, ``The Rise of AI Companions: How Human-Chatbot Relationships Influence Well-Being,'' \emph{arXiv preprint arXiv:2506.12605}, 2025.

\bibitem{wang2025understanding}
Y.~Wang, Y.~Wang, K.~Crace, and Y.~Zhang, ``Understanding Attitudes and Trust of Generative AI Chatbots for Social Anxiety Support,'' in \emph{Proceedings of the 2025 CHI Conference on Human Factors in Computing Systems}, 2025, pp. 1--21.

\bibitem{lee2025large}
\BIBentryALTinterwordspacing
O.~Lee and K.~Joseph, ``A large-scale analysis of public-facing, community-built chatbots on Character.AI,'' 2025. [Online]. Available: \url{https://arxiv.org/abs/2505.13354}
\BIBentrySTDinterwordspacing

\bibitem{karimova2025ethical}
G.~Z. Karimova, ``Ethical Foundations of AI with Personality,'' in \emph{Humanizing AI with Personality}.\hskip 1em plus 0.5em minus 0.4em\relax Springer, 2025, pp. 19--31.

\bibitem{filatov2024development}
E.~Filatov, ``Development of students’ foreign language communicative skills based on the character. ai web application,'' \emph{Tambov University Review. Series: Humanities}, vol.~29, no.~5, pp. 1248--1260, 2024.

\bibitem{guan2025unpacking}
H.~Guan, J.~Jamieson, G.~Gao, and N.~Yamashita, ``Unpacking Negative Feelings and Perceptual Gaps About Social Interactions with Conversational AI,'' in \emph{Proceedings of the Extended Abstracts of the CHI Conference on Human Factors in Computing Systems}, 2025, pp. 1--8.

\bibitem{ragab2024trust}
A.~Ragab, M.~Mannan, and A.~Youssef, ```Trust Me Over My Privacy Policy': Privacy Discrepancies in Romantic AI Chatbot Apps,'' in \emph{2024 IEEE European Symposium on Security and Privacy Workshops (EuroS\&PW)}.\hskip 1em plus 0.5em minus 0.4em\relax IEEE, 2024, pp. 484--495.

\bibitem{qwen3guard}
\BIBentryALTinterwordspacing
Q.~Team, ``Qwen3Guard Technical Report,'' 2025. [Online]. Available: \url{http://arxiv.org/abs/2510.14276}
\BIBentrySTDinterwordspacing

\bibitem{wildguard2024}
\BIBentryALTinterwordspacing
S.~Han, K.~Rao, A.~Ettinger, L.~Jiang, B.~Y. Lin, N.~Lambert, Y.~Choi, and N.~Dziri, ``WildGuard: Open One-Stop Moderation Tools for Safety Risks, Jailbreaks, and Refusals of LLMs,'' 2024. [Online]. Available: \url{https://arxiv.org/abs/2406.18495}
\BIBentrySTDinterwordspacing

\end{thebibliography}
